\documentclass[aps,pra,onecolumn,nobibnotes,superscriptaddress,nofootinbib]{revtex4-2}
\usepackage{graphicx} 
\usepackage{caption,graphics,physics}
\usepackage{subcaption}
\usepackage{amsmath,xcolor,bm}
\usepackage{qcircuit}
\usepackage{mathtools,amsmath,amsthm,amsfonts,amssymb,amscd}
\usepackage{hyperref}
\usepackage[ruled,vlined]{algorithm2e}
\usepackage{algpseudocode}

\usepackage[capitalize]{cleveref}

\hypersetup{
    colorlinks=true,
    linkcolor=red,
    filecolor=magenta,      
    urlcolor=cyan,
}
\usepackage{geometry}

\def\polylog{\mathrm{polylog}}

\newtheorem{theorem}{Theorem}

\newtheorem{lemma}[theorem]{Lemma}

\newtheorem{remark}[theorem]{Remark}

\newtheorem{assumption}[theorem]{Assumption}
\newtheorem{definition}[theorem]{Definition}

\newtheorem{problem}{Problem}
\newcommand{\bs}{\boldsymbol}

\newcommand{\ch}{\mathcal{H}}
\newcommand{\co}{\mathcal{O}}

\Crefname{appsec}{Appendix}{Appendices}

\begin{document}
%
\title{ Implementation of the Density-functional Theory on Quantum Computers with Linear Scaling with respect to the Number of Atoms}

\author{Taehee Ko}
\email{tuk351@psu.edu}
\affiliation{Department of Mathematics, Pennsylvania State University}

\author{Xiantao Li}%
\email{xiantao.li@psu.edu.}
\affiliation{Department of Mathematics, Pennsylvania State University}

\author{Chunhao Wang}%
\email{cwang@psu.edu}
\affiliation{Department of Computer Science and Engineering, Pennsylvania State University}

\begin{abstract}
   Density-functional theory (DFT) has revolutionized computer simulations in chemistry and material science.  A faithful implementation of the theory requires self-consistent calculations. However, this effort involves repeatedly diagonalizing the Hamiltonian, for which a classical algorithm typically requires a computational complexity that scales cubically with respect to the number of \emph{electrons}. This limits DFT's applicability to large-scale problems with complex chemical environments and microstructures. This article presents a  quantum algorithm that has a linear scaling with respect to the number of \emph{atoms}, which is much smaller than the number of electrons. Our algorithm leverages the quantum singular value transformation (QSVT) to generate a quantum circuit to encode the density-matrix, and an estimation method for computing the output electron density. In addition, we present a randomized block coordinate fixed-point method to accelerate the self-consistent field calculations by reducing the number of components of the electron density that needs to be estimated. 
   The proposed framework is accompanied by a rigorous error analysis that quantifies the function approximation error, the statistical fluctuation, and the iteration complexity.  In particular, the analysis of our self-consistent iterations takes into account the measurement noise from the quantum circuit.
These advancements offer a promising avenue for tackling large-scale DFT problems, enabling simulations of complex systems that were previously computationally infeasible. 
\end{abstract}
\maketitle

\section{Introduction}

One of the breakthroughs in computational chemistry is the development of the density-functional theory (DFT) \cite{hohenberg1964inhomogeneous}, which led to the Nobel Prize in Chemistry in 1998. The theory is founded on the observation that the electronic structures are fully determined by the underlying electron density $ {n}(\bm r); \bm r \in \mathbb{R}^3 \to [0, +\infty )$  which, thanks to the remarkable work of Kohn and Sham \cite{Kohn1965}, can be represented through an auxiliary system of non-interacting electrons with an effective Kohn-Sham Hamiltonian $\ch$. The electron-electron interactions are captured by an exchange-correlation energy functional of ${n}(\bm r)$, which is part of the Hamiltonian operator. In the implementation, the electron density is computed self-consistently to meet the self-consistent field (SCF) requirement \cite{parr1995density}, which mathematically corresponds to a fixed-point problem \cite{lin2013elliptic,toth_convergence_2015}. Meanwhile, direct computation amounts to calculating many eigenvalues of a large-dimensional matrix, for which the computational cost typically scales cubically with the dimension \cite{zhou2006self}. Such scaling has been the major limiting factor for large-scale DFT calculations. The most expensive component in computing the new electron density is the eigenvalue calculations.   Due to such cubic scaling, large-scale DFT calculations are still an outstanding challenge in connecting electron structure to macroscale material properties, e.g., perovskite materials \cite{hautier2010finding}, high-entropy alloys \cite{chen2021simultaneously}, bi-layer two-dimensional materials with small twist angle \cite{yoo2019atomic}, and biomaterials \cite{elstner2000self}. 

The purpose of this paper is to demonstrate a quantum speedup with a new quantum algorithm that leverages many unique capabilities of quantum computing devices.  
Instead of explicitly computing the eigenvalues and eigenvectors, we apply an eigenvalue transformation and construct a quantum circuit for the density-matrix in DFT. The electron density is then extracted from the diagonals of the density-matrix. We will show that the gate and query complexity of the algorithm only scales linearly with respect to the dimension of the problem, which we will compress down to be proportional to the number of atoms. In addition, we propose an efficient self-consistent iteration algorithm, where only some components of the electron density are updated.  We provide theoretical analysis for the convergence of the iteration methods. Our numerical results indicate that the overall complexity can be well below the theoretical bound.

The paper is organized as follows. In the remainder of this introduction, we provide informal problem statements and summarize our main results, followed by discussions of related works.  In \cref{sec: setup}, we detail the problem setup and highlight the computational aspects, including the spatial discretization and SCF iterations. Our quantum algorithms will be presented in \cref{sec: q-alg}, together with error estimates and complexity bounds.   We show some numerical results in \cref{sec:num}.

\subsection{Problem Statements and Summary of Results }

The self-consistent field (SCF) in DFT asserts that the electron density $n(\bm r) $ that enters the effective Kohn-Sham Hamiltonian $\ch$ has to be the same as the electron density determined from the eigenvalues and eigenvectors of $\ch $. This is often achieved by iterations, $\bm n \to F(\bm n)$, with $\bm n$ and $F(\bm n)$ being respectively the input and updated density represented at grid points (see the precise definition in \cref{fixed-point-problem2}). The components of $F(\bm n)$ can be linked to a density-matrix, expressed as the diagonals of a matrix function, $f(\ch )$, with $f$ being the Fermi-Dirac function $f(x)=\left(1+\exp \beta (x-\mu) \right)^{-1}$ at finite temperature ($\beta =\co(1)$). The first problem addresses the computation of $F(\bm n)$.

\begin{problem}[Updating the electron density]
Assume that the effective Hamiltonian $\ch $ is approximated by a Hermitian matrix $H$ with sparsity $s$ on a set of grid points. Suppose we are given an oracle to access $H$ and its nonzero elements (See \cref{eq:os,eq:oh}). 
   Determine an estimate for the updated electron density $\hat{F}(\bm n)\in \mathbb{R}^{N_I}$ at $N_I$ grid points, 
   such that $\norm{\hat{F}(\bm n) - {F}(\bm n) }< \epsilon$. 
\end{problem}
\begin{theorem}[Informal version of \cref{prop:Fa}] 
     There is a quantum algorithm that outputs an approximate electron density $\hat{F}(\bm n)$ such that $\norm{\hat{F}(\bm n) - {F}(\bm n) }< \epsilon$ with probability at least $1-\delta $. Under the assumptions above, the algorithm involves a query complexity $\widetilde{\mathcal{O}}\big(\frac{s N_I}{\epsilon}\big)$.\footnote{We use $\widetilde{\mathcal{O}}$ to neglect poly-logarithmic factors.} 
\end{theorem}

The updated density $F(\bm n)$ can be used as the input density at the next iteration, and the iterations continue until the two densities coincide. In the limit, the electron density converges to the ground state density $\bm n_*$, i.e., $n_* = F(\bm n_*)$.

\begin{problem}[Determining the \emph{ground state} electron density]
   Given an input electron density $n_0(\bm r)$, such that $\norm{n_*(\bm r) - n_0(\bm r)}< \gamma,$ with sufficiently small $\gamma$. 
   Determine an estimate $\hat{n}(\bm r)$  such that $\norm{n_*(\bm r) - \hat{n}(\bm r)}< \epsilon.$ 
\end{problem}

\begin{theorem}[Informal version of \cref{thm: full}] 
Assume that the effective Hamiltonian $\ch $ is approximated by $H$ on a set of grid points and $H$ has sparsity $s$.  Suppose we are given an oracle to access $H$ and its nonzero elements (See \cref{eq:os,eq:oh}). 
     There is a hybrid quantum-classical algorithm that outputs an approximate ground state electron density $\hat{n}(\bm r)$ such that $\norm{n_*(\bm r) - \hat{n}(\bm r)}< \epsilon.$  Neglecting logarithmic factors, and under the assumptions above, the algorithm involves  $\widetilde{\mathcal{O}}\big(\frac{sN_I}{\epsilon}  \big) $ queries to the Hamiltonian matrix $H$.
\end{theorem}

Overall, our approach is a quantum-classical hybrid algorithm, where the quantum algorithm produces the density-matrix $f(H)$ (see also $\Gamma$  in \cref{mat-gam1}), while the classical algorithm employs an SCF iteration to provide updated values of the electron density to reprogram the quantum algorithm by updating $H$ at the next step. 
 As a result, the updated density is subject to measurement noise. This gives rise to a \emph{stochastic} SCF problem. In addition to a straightforward application of a mixing scheme \cite{bowler2000efficient,haydock1972electronic}, we propose a random coordinate method,  in which, for each fixed-point iteration (see \cref{fixed-point-problem2}), one only chooses to update some randomly selected components of $\hat{F}(\bm n)$, rather than computing all the components. This significantly reduces the number of measurements needed at each iteration. The theoretical analysis shows that the new method has a comparable convergence rate as the full coordinate method that computes all components at each iteration step, 
 and numerical tests suggest that the new method can be significantly more efficient overall.  Namely, the complexity can be sublinear in $N_I$, the number of grid points.

Compared to classical algorithms, the hybrid algorithm has far better scaling in terms of the number of electrons.  This quantum speed-up offers a promising opportunity for treating large-scale DFT problems, and it has the exciting potential to lead to accelerated discoveries enabled by DFT.

\subsection{Related work}

\noindent{\bf Classical algorithms for computing the updated electron density} 
In classical computing, the most expensive part of typical DFT implementations to compute the electron density is the step of solving the Kohn-Sham equations, which is equivalent to finding eigenpairs corresponding to the  Hamiltonian matrix. Many efficient techniques have been proposed over the last two decades, such as polynomial filtering methods \cite{bekas2008computation,zhou2006self,liou2020scalable}, direct energy minimization \cite{vecharynski2015projected,wen2016trace} and spectrum slicing type methods \cite{schofield2012spectrum,li2016thick}. For numerical implementations of DFT, the readers are referred to \cite{martin2004@book,lin2019mathematical,lin2019numerical}, and a large collection of software packages \cite{hafner2008ab,yang2009kssolv,liou2020scalable,seifert2012density,gale2011siesta,sharma2018calculation,motamarri2020dft}.  The complexity of these algorithms typically scales cubically with the number of electrons:  $\mathcal{O}(N_e^3)$. There are several classical algorithms for electron structure calculations that exhibit linear  (potentially sublinear scaling) with respect to the number of electrons \cite{soler2002siesta,cleri1993tight,goedecker1999linear,garcia2007sub,gavini2007quasi}. Although such complexity is particularly attractive for large-scale problems in material science and chemistry, these methods do not strictly satisfy self-consistency. Therefore it is difficult to quantify the error when compared to the true ground-state electron density.

\medskip

\noindent{\bf Quantum algorithms for electron structure calculations.}
Studying electron structures on quantum computers has been envisioned to be one of the first few applications of quantum computing. Many algorithms have been proposed to push this effort forward based on the many-body Schr\"odinger equation \cite{abrams1997simulation,tesch2001applying,babbush2014adiabatic,hastings2014improving,aspuru2005simulated,o2016scalable,babbush2018low,yoo2019atomic,lin2022heisenberg}. But the ability of these algorithms to handle large-scale problems, e.g., those in \cite{hautier2010finding,chen2021simultaneously,yoo2019atomic,elstner2000self} that are of direct practical interest, has not been demonstrated. We expect that even for fault-tolerance quantum computers, mean-field models, such as DFT, will still be an important alternative. Meanwhile,
due to its origin in many-body quantum physics, DFT has been studied in the context of quantum computing \cite{baker2020density,gaitan2009density,senjean2023toward}. The work of Baker and Poulin \cite{baker2020density} attempts to compute the Kohn-Sham potential with quantum computing; Gaitan and Nori \cite{gaitan2009density} demonstrated that DFT can be formulated for a quantum system consisting of qubits; Senjean et al. \cite{senjean2023toward} showed how the Kohn-Sham Hamiltonian, constructed based on an auxiliary non-interacting system, can be mapped to an interacting Hamiltonian on quantum computers. None of the aforementioned works, however, solves the self-consistent DFT model directly on quantum computers. More importantly, precise error estimates and the overall computational complexity were not addressed. To the authors' knowledge, this paper is the first attempt to faithfully implement DFT on quantum computers with rigorous complexity estimates and direct comparison of the complexity with classical algorithms. 

\medskip 

\noindent{\bf Classical algorithms for the self-consistent iterations.} The most widely used algorithm for the SCF iterations is the mixing schemes \cite{bowler2000efficient,haydock1972electronic,lin2013elliptic}, which can be proved to be at least linearly convergent \cite{lin2013elliptic,toth_convergence_2015,toth_local_2017}.   Therefore, the complexity, in terms of the number of iterations until convergence, involves a $\log \frac{1}{\epsilon}$ scaling. Our hybrid algorithm, which uses a block encoding \cite{GSLW19} of the density-matrix ($\Gamma$ in \cref{mat-gam1}) and the amplitude amplification \cite{brassard2000amplitude} to estimate the electron density, has the same complexity in the fixed-point iterations. Motivated by the remarkable success of stochastic approximation methods \cite{robbins1951stochastic,wolfowitz1952stochastic,chung1954stochastic,nemirovski2009robust,ghadimi2013stochastic} in the optimization of large-scale machine learning models \cite{bottou2018optimization}, 
the authors Ko and Li \cite{ko2023stochastic} have recently developed a classical algorithm to carry out the self-consistent iterations stochastically, so that each iteration has a linear scaling complexity.  The core of the algorithms in both \cite{ko2023stochastic} and \cite{baer2013self} is a randomized numerical method called the trace estimator~\cite{bekas2007estimator,meyer2021hutch++,lin2017randomized,persson2022improved,hallman2022multilevel,avron2011randomized,lin2016approximating}. Such an approach was applied in computational chemistry~\cite{baer2013self} for the DFT, albeit without rigorous error bounds. For the computational complexity, the total number of iterations has a polynomial dependence on $\epsilon.$ In addition, each step of the iteration still requires a diagonalization of a small matrix. These limitations will be removed in the present framework, by using quantum computing algorithms.

\noindent{\bf Random coordinate methods in machine-learning and reinforcement-learning.}  Coordinate-wise iterative algorithms are the state of the art for many large-scale problems, due to their simplicity, low cost per iteration, and overall efficiency. Many variants have been developed and improved convergence than the full coordinate counterparts has been demonstrated for optimization problems, see ~\cite{nesterov2012efficiency,nesterov2017efficiency,nutini2015coordinate,lin2014accelerated,saha2013nonasymptotic,wright2015coordinate,karimireddy2019efficient,chen2023global} and the references therein. What is closely related to the current work is the idea of the coordinate-wise update rule has been used in the context of fixed-point problems~\cite{chow2017cyclic,iiduka2019stochastic,combettes2015stochastic,tsitsiklis1994asynchronous,peng2016arock} but less explored than in the optimization tasks. Among these works, the asynchronous coordinate update rules in \cite{tsitsiklis1994asynchronous,peng2016arock} have a resemblance to the proposed algorithm, as those rules randomly update a portion of components of fixed-point mappings at each iteration. But there are fundamental differences in our approach. For example, the focus of \cite{tsitsiklis1994asynchronous} is on the Q-learning in the context of reinforcement learning, which requires more restrictive assumptions due to its complicated problem setup. In \cite{peng2016arock}, the authors consider situations where the estimation of a component of a mapping is exact while in our setting, it involves random noise due to the nature of quantum measurement. Besides, we provide convergence analysis of the block coordinate case which was not analyzed in \cite{peng2016arock}.

\section{Problem Setup}\label{sec: setup}

\paragraph*{Notations.} We use bold fonts for vectors, e.g., $\bm r$, and the entries will be labeled in parenthesis, e.g., $\bm r(j)$ being the $j$th entry of $\bm r$. $D \subset \mathbb{R}^3$ will be used to denote the physical domain. $n(\bm r): D \to [0,+\infty) $ is a function representing the electron density. Here is a summary of the notations that will be used in this section. $N_e$ and $N_a$ are respectively the number of electrons and the number of atoms. In the numerical discretization, $D_\delta$ is a set of grid points in $D$, with $N_g$ being the number of grid points, which is often comparable to $N_e$. We choose $N_g=2^m$ to map functions defined at the grid points to quantum states in a $m$-qubit system. 
$H\in \mathbb{R}^{N_g\times N_g}$ refers to the Hamiltonian represented at the grid points. 
Meanwhile,  $D_\Delta$ is a set of coarse grid points in $D$ and $N_I=\abs{D_\Delta} $, which is comparable to $N_a.$

\paragraph*{The DFT formulations.} DFT is formulated to find the ground state energy of a system by solving the following eigenvalue problem \cite{Kohn1965},
\begin{equation}\label{eig}
\begin{split}
    &\ch[n]  \ket{\psi_j }= E_j \ket{ \psi_j}, \\
    &\ch[n]  = -\frac{\nabla^2}{2}+ V[n](\bm r), \\
    &V[n](\bm r):= V_H[n](\bm r) + V_\text{xc}[n](\bm r) + V_\text{ext},\\
    &\int\psi_i(\bs{r})^*\psi_j(\bs{r})d\bs{r}=\delta_{ij},
\end{split}
\end{equation}where  $E_j$'s are the Kohn-Sham eigenvalues and $\psi_j$'s are the Kohn-Sham wavefunctions. The notation $[\cdot]$ indicates a dependence on the function $n(\bm r)$ of a functional.  The first term in $H$ is the one-electron kinetic energy.  $V_H[n](\bm r)$ is the Hartree potential, which is a functional of $n$. More precisely,  this potential can be obtained by solving the Poisson equation \cite{ghosh2017sparc}, 
\begin{equation}\label{poissoneq}
    -\frac{1}{4\pi}\nabla^2V_H(\bm r) = n(\bm r) + b.
\end{equation}In the above equation, $b$ comes from the pseudocharges from the nuclei 
and other possible charge corrections. The second term $V_\text{xc}[n]$ embodies the electron-electron interactions. This function is universal and it has been parameterized in function forms that are easily implementable, e.g., \cite{perdew1992accurate,marques2012libxc}. Finally, the external potential energy $V_\text{ext}$ accounts for the interaction between electrons and nuclei.

From the eigenvalue problem in \cref{eig}, the finite temperature density matrix operator, which is known as the first-order density matrix \cite{parr1995density}, is defined as
\begin{equation}\label{mat-gam0}
    \rho_1(\bm r, \bm r') \coloneqq f(\ch ),
\end{equation}where $f$ is the Fermi-Dirac function with the inverse temperature $\beta$ and the chemical potential $\mu$,
\begin{equation}\label{fdfunc}
    f(x) = \frac{1}{1+ \exp (\beta (x - \mu))  }.
\end{equation}

The computation of $\rho_1$ requires a given chemical potential $\mu$. 
In the case where the number of electrons $N_e$ is fixed,  $\mu$ is chosen such that
\begin{equation}\label{constraint}
    \sum_jf(E_j) =  N_e,
\end{equation}where $f(E_j)$ is referred to as occupation numbers and $N_e$ is the number of electrons. Here we neglect the spin orbitals for simplicity.

The eigenvalue problem in \cref{eig} provides an implicit representation of the electron density. In particular, the electron density $n(\bm r)$ defined by  
\begin{equation}\label{eden}
     n(\bm r) \coloneqq \sum_{j} f(E_j) \abs{\psi_j(\bm r)}^2,
\end{equation}
which in turn determines $V_H$ and $V_\text{xc}$ in \cref{eig} and therefore the Hamiltonian $\ch[n] $. 

Using the matrix function notation, we can express the problem of determining the electron density as the following fixed-point problem,
\begin{equation}\label{fixed-point-problem1}
    n(\bm r) = \bra{\bm r}f\big(\ch [n]\big)\ket{\bm r}, 
\end{equation}
where $\ch[n]$ has been defined in \cref{eig}.

\begin{remark}
It is important to point out that although it has been customary in the DFT literature \cite{parr1995density,martin2004@book} to refer to $f(\ch )$ as a density-matrix, it may not have trace one, which is usually required for a density-matrix in quantum information.     
\end{remark}

\subsection{Real-space Discretization}\label{sec: real space discretization}
To solve \cref{eig} in a computation, we assume that the Hamiltonian operator is properly discretized in a three-dimensional domain $D$ by a finite-difference method \cite{beck_real-space_2000} with grid size $\delta$. We denote $H\in \mathbb{C}^{M\times M}$ to be the Hamiltonian matrix; $M=2^m$, so that it can be directly mapped to the Hilbert space associated with a quantum circuit with $m$ qubits. 

Within the discretization, the electron density at the grid points is expressed as a vector $\bm n.$
Following the Hamiltonian operator in \cref{eig}, we can express the matrix $H$ as follows,
\begin{equation}
    H(\bm n) = -\frac{1}{2} \nabla_\delta^2 +  V_\delta(\bm n),
\end{equation}
where $ \nabla_\delta^2$ is a finite-difference approximation of the kinetic energy operator. $V_\delta$, which enters the Hamiltonian through the diagonals, is the potential evaluated at the grid points and it collects all the potential terms in the Hamiltonian operator.

In terms of the matrix $H$ from the finite-difference approximation, we can define the density-matrix $\Gamma \in \mathbb{C}^{2^m\times 2^m}$ that is similar to \cref{mat-gam0},
\begin{equation}\label{mat-gam1}
    \Gamma = f({H}). 
\end{equation}

Similarly, we generalize the continuous fixed-point problem in \cref{fixed-point-problem1} to a discrete one,
\begin{equation}\label{fixed-point-problem3}
    \bm n =  f\left({H(\bm n)}\right). 
\end{equation}

 Real-space discretizations usually lead to sparse Hamiltonian matrix $H$. Therefore, we assume that in the following quantum algorithms, the Hamiltonian $H$ is $s$-sparse, in the sense that there are at most $s$ nonzero entries in each row/column. $s$ depends on the choice of the finite difference methods. For example, if the second-order central difference method is used, then $s=7$, and a fourth-order method would give the sparsity $s=13$ (see \cite{ghosh2017sparc}).

\begin{figure}[htp]
    \centering
    \includegraphics[scale=0.2]{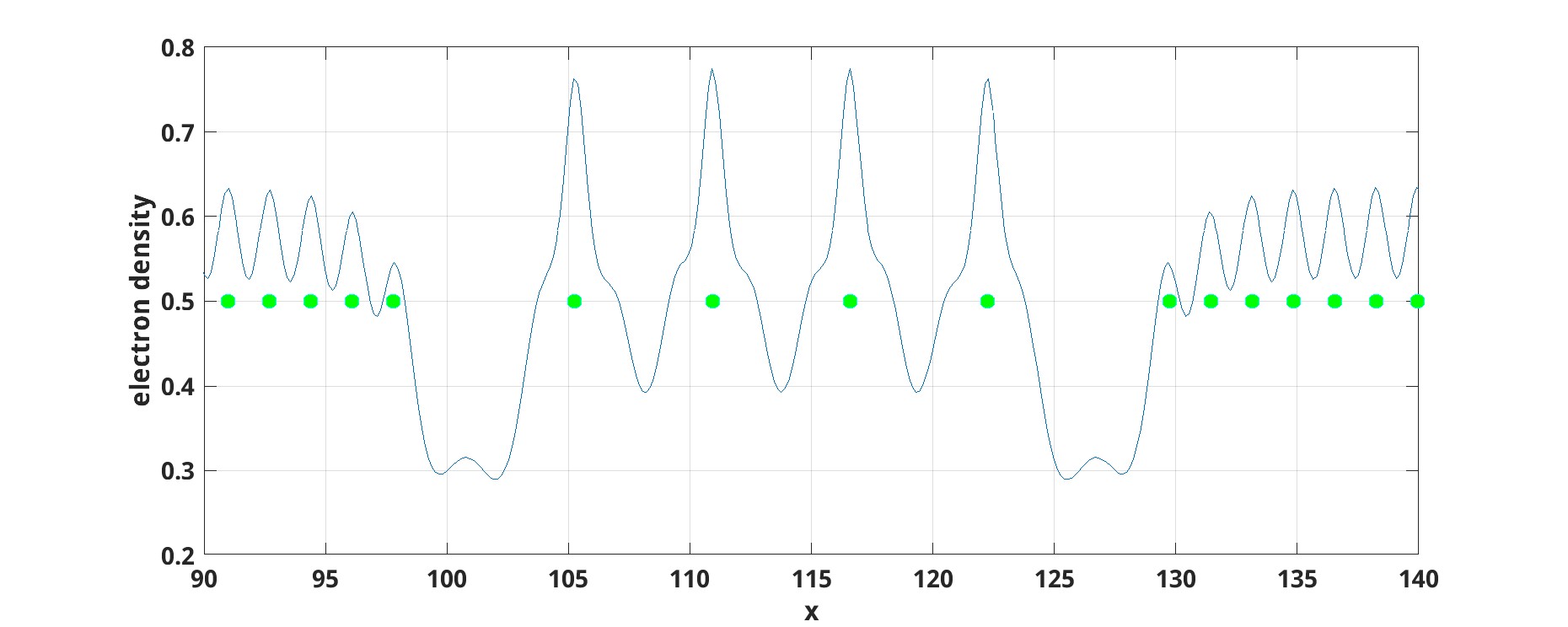}
    \caption{A quick illustration of the ground state electron density of a one-dimensional Lithium-Hydrogen chain from \cite{ke2007role}. Also shown are the atoms (6 Li atoms in the middle and H atoms on the two sides).   }
    \label{fig:eden}
\end{figure}

To further make the algorithms more practical, we discuss how the problem size $M$ can be reduced by means of interpolation techniques.  As observed in \cref{fig:eden}, the electron density typically forms a smooth function peaked around $N_a$ atoms. From this observation, we can take $N_I$ interpolation points, with $N_I$ comparable to $N_a$, so that the original electron density on the entire $M$ grid points can be effectively represented without compromising the accuracy. The electron density will be used to compute the potential $V_\delta$ at the interpolation points and interpolated to reconstruct the Hamiltonian.   More precisely,
 assuming that we have the potential $V_\delta(\bm r)$ obtained only at $N_I$ interpolation points ($N_I\ll M$), we construct its interpolation such that for any $k\in[M]$, 

\begin{equation}\label{interp}
 V_\delta(\bm r_k) \approx  \sum_{\bm r \in D_\Delta} V_\delta(\bm r) \mathcal{N} (\bm r_k - \bm r),
\end{equation}
where $D_{
\Delta}$ is the set of the interpolation points (namely $|D_{\Delta}|=N_I$) and the function $\mathcal{N}$ are the shape functions. In classical algorithms for DFT,  
this interpolation is part of the multigrid scheme to compute $V_H$ in the Poisson equation (\cref{poissoneq}) in DFT \cite{merrick1995multigrid}. Besides, $V_\text{xc}$ can be simply determined from the interpolated electron density with explicit functional evaluation (e.g. the local density approximation \cite{perdew1981local}). Therefore, we can efficiently construct the Hamiltonian matrix $H\in\mathbb{C}^{M\times M}$ from the $N_I$-dimensional interpolated electron density. As we will discuss in \cref{sec: preliminaries}, this interpolation leads to an appreciable reduction of quantum random-access memory (QRAM) storage, thereby yielding an $\tilde{\mathcal{O}}(N_I)$ scaling implementation of the Hamiltonian on quantum hardware. 

With this interpolation, we reduce the fixed-point problem in \cref{fixed-point-problem3} to one that is defined on $D_\Delta$,
\begin{equation}\label{fixed-point-problem2}
    \bm n(j) =  f\left({H(\bm n)}\right)(j,j) =: F(\bm n)(j), j \in [N_I]. 
\end{equation}
Namely, each diagonal must match with the input electron density at the grid point $\bm r_j \in D_\Delta$. With a slight abuse of the notations, we will still denote this reduced fixed-point problem as $\bm n = F(\bm n)$.

\subsection{Self-consistent iterations}

Like many mean-field theories in quantum chemistry, \cref{eden,eig} have to be solved self-consistently. 
At the level of numerical discretization, this is manifested as the fixed-point problems in \cref{fixed-point-problem2}.

\begin{figure}[thb]
    \centering
    \includegraphics[scale=0.23]{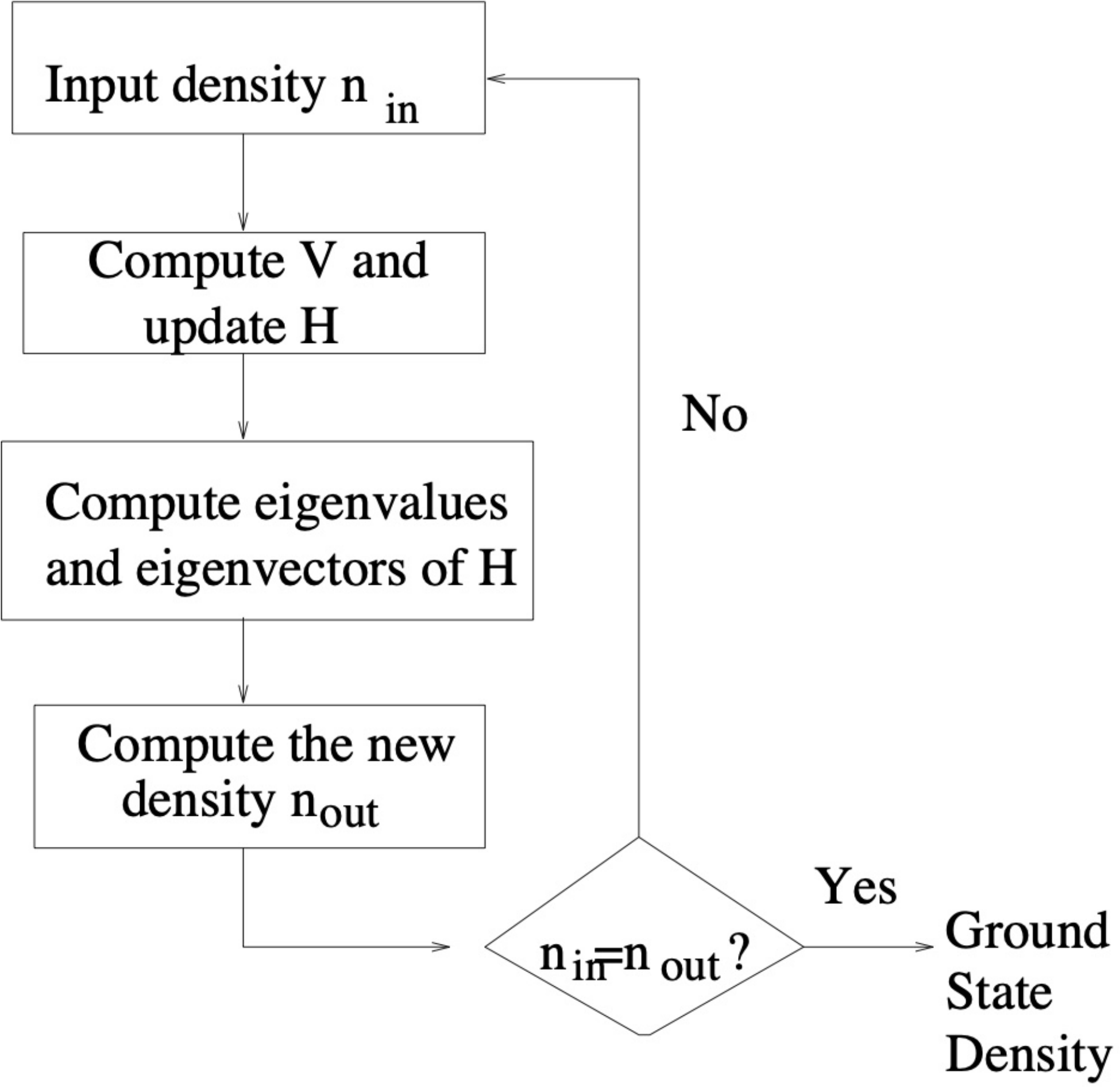}
    \caption{An illustration of the SCF procedure in DFT.}
    \label{fig:scf}
\end{figure}

In terms of implementation, the nonlinear mapping $F$ is implicitly determined by the procedure of obtaining the output from input $\bm n$ within the SCF iteration as shown in \cref{fig:scf}. 
A simple procedure to obtain a fixed point is to apply iterations $ \bm n_{k+1}= F( \bm n_k)$ repeatedly until convergence. To guarantee and speed up convergence, mixing schemes are typically applied in practice, such as simple mixing and Pulay mixing.  This will be explained in \cref{hybrid}.

\section{Quantum Algorithms}\label{sec: q-alg}

\subsection{Preliminaries}\label{sec: preliminaries}

As pointed out in the previous section, the matrix $H$ from the real-space discretization is usually sparse. The sparsity implies that the matrix is \emph{efficiently row/column computable}. To access $H$, we assume we have access to a procedure ${O}_S$  that can perform the following mapping: 
  \begin{align}
  \label{eq:os}
    {O}_S: \ket{i}\ket{k} \mapsto \ket{i}\ket{r_{i_k}},
  \end{align}
  where $r_{i_k}$ is the $k$-th nonzero entry of the $i$-th row of $A$. In addition, ${O}_H$ can also perform the following mapping:
  \begin{align}
  \label{eq:oh}
    {O}_H: \ket{i}\ket{j}\ket{0} \mapsto \ket{i}\ket{j}\ket{H(i, j)}.
  \end{align}

One key ingredient of our quantum algorithm is block encoding. We say that $U_A$ is an $(\alpha,a,\epsilon)$-block-encoding of $A$ if $U_A$ is a $(m + a)$-qubit unitary, and
\begin{align}\label{be-H}
  \norm{A - \alpha  (\bra*{0^{\otimes a}}\otimes I)U_A(\ket*{0^{\otimes a}}\otimes I)}_2 \leq \epsilon.
\end{align}
Intuitively, the block encoding constructs a unitary with the upper-left block being proportional to $H,$ 
\[  U_A = 
  \begin{pmatrix}
    A & \vdot \\
   \vdot & \vdot
  \end{pmatrix}.\]
To implement ${O}_S$ and ${O}_H$ efficiently, we use 
 the interpolation in \cref{interp} to generate the electron density in \cref{fixed-point-problem1} approximately. For this, we need to store $\co(N_I)$ parameters in QRAM in order to update the diagonals of $H$  input oracle. The gate complexity for implementing such QRAM is ${O}(N_I)$. Moreover, the circuit depth of QRAM is $\co(\log N_I)$ \cite{nielsen2011quantum}. The input oracles ${O}_S$ and ${O}_H$ for $H$ can be implemented as a procedure that reads data in the QRAM.

\begin{figure}[thb]
    \centering
    \includegraphics[scale=0.5]{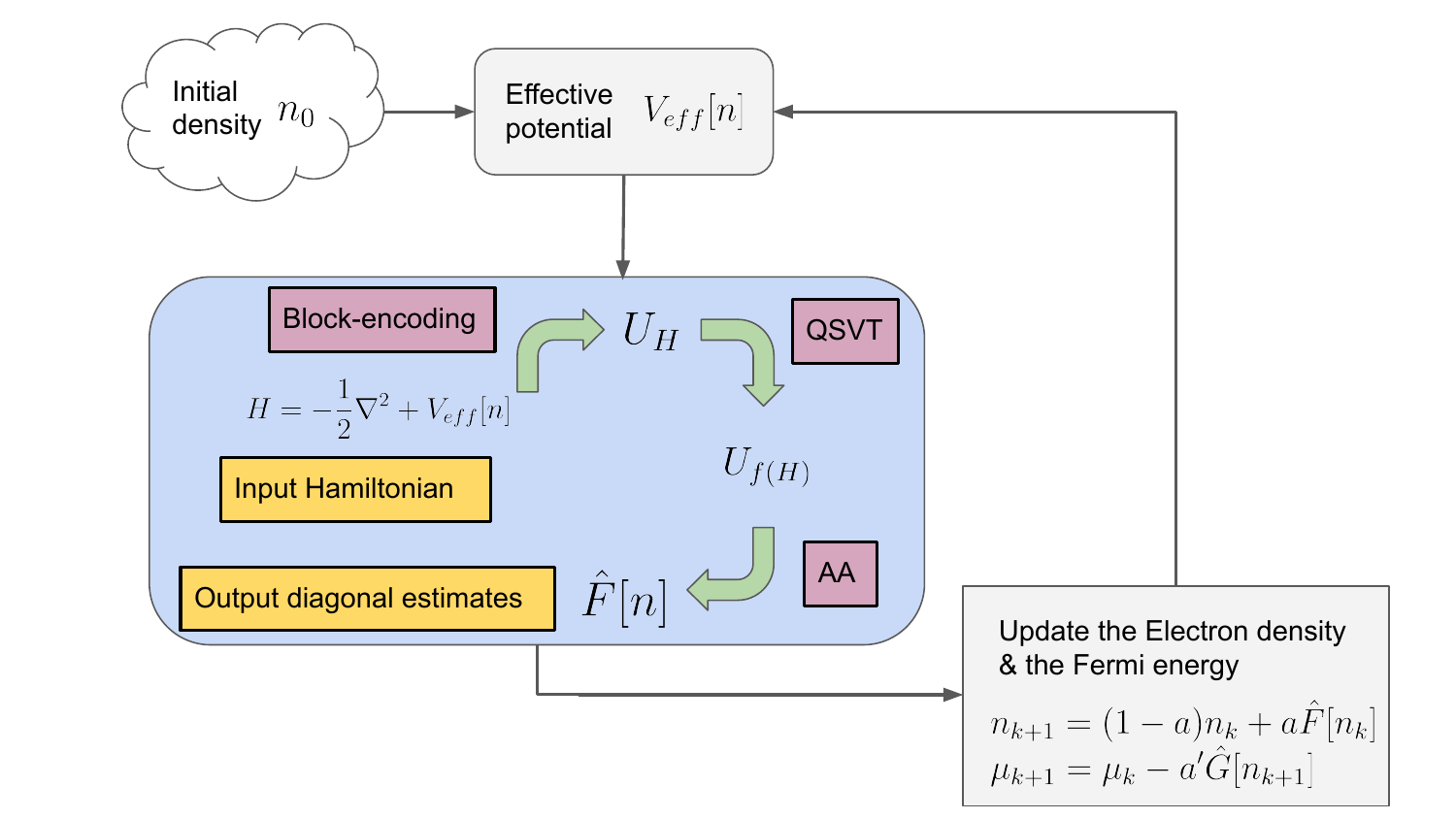}
    \caption{An illustration of the hybrid algorithm.}
    \label{fig:my_label}
\end{figure}

In the next three sections, we will present our quantum algorithm. We first outline a high-level description of the algorithm in \cref{fig:my_label}, which consists of a quantum singular value transformation (QSVT) to construct a quantum circuit for the density-matrix, an amplitude amplification (AA) to estimate the updated electron density and a classical fixed-point iteration to provide the electron density (and chemical potential if $N_e$ is given )  for the next iteration.

\subsection{Preparing the density-matrix using quantum singular value transformation} 

Since $H$ is Hermitian, one can use the spectral map and approximate the density-matrix in \cref{mat-gam1} by polynomial approximations of the Fermi-Dirac function (\cref{fdfunc}). For the error analysis, we use a result from function approximation \cite[Theorem 8.2]{trefethen2019approximation}, which is restated as follows, 
\begin{lemma}\label{lemma: approximate Fermi-dirac}For any analytic function $f$ such that can be analytically extended to an Berstein ellipse $B_r$ with some $r>1$,  there exists a polynomial $p_\ell$ of degree $\ell$ such that
    \begin{equation}\label{pell}
       \max_{x\in [-1,1]} \abs{{f(x)} - p_\ell(x) } \leq 2\sup_{z\in B_{r}}|{f(z)}|\cdot \frac{r^{-\ell}}{r - 1}.
    \end{equation} 
\end{lemma}
To apply this technique to the density-matrix in \cref{mat-gam1}, we rescale the Hamiltonian matrix as follows 
\begin{equation}
    f(H) = \left(1+\exp\left(\beta\left(\frac{\lambda_{+}+\lambda_{-}}{2}-\mu\right)\right)\exp(\hat{\beta}\tilde{H})\right)^{-1}, 
\end{equation}where
\begin{equation}\label{eq: scaled quantities}
    \hat{\beta} \coloneqq \frac{\lambda_{+}-\lambda_{-}}{2}\beta,\quad \tilde{H} \coloneqq  \frac{2}{\lambda_{+}-\lambda_{-}}\left(H-\frac{\lambda_{+}+\lambda_{-}}{2} I  \right). 
\end{equation}Here $\lambda_{-}$ and $\lambda_{+}$ are some lower and upper bounds of the eigenvalues of $H$.  The scaling is simply to map the eigenvalues of $H$ to the interval $[-1,1].$
 One way to roughly estimate an upper bound is to apply Gershgorin's Circle Theorem.  A tighter upper bound can be efficiently obtained by running only a few steps of the Lanczos algorithm, as pointed out in ~\cite{zhou2006self}, which will take $\mathcal{O}(s M)$ operations. Once we get an estimate of an upper bound of $\lambda_{+}$, we can also obtain a lower bound of $\lambda_{-}$ in a similar manner after shifting $H$ properly.

Noticing that $\sigma(\tilde{H})\subset [-1,1]$, we can apply \cref{lemma: approximate Fermi-dirac} for the polynomial approximation of the density matrix. The following lemma, as in \cite[Remark 4.8]{ko2023stochastic}, shows that the quality of approximation depends on a given temperature. With a slight abuse of notations, we will continue to use $H$ as the scaled Hamiltonian.  

\begin{lemma}\label{lem:ell}
    For a given inverse temperature $\beta$, the degree of the Chebyshev expansion to approximate the $f(H)$, up to a precision $\epsilon$, requires at least,
    \begin{equation}
        \ell = {\Theta}\left(\log_{r}\frac{1}{\epsilon}\right).
    \end{equation}
Here the constant $r$ satisfies that  $r\in (1, \frac{c(\beta)+\sqrt{c(\beta)^2+4}}{2})$ with $c(\beta) = \frac{4\pi}{\lambda_+-\lambda_-}\frac{1}{\beta}$. 
\end{lemma}\label{low-temp}

\begin{remark}
    We observe that at the low temperature where $\hat{\beta}\gg 1$, we have $r\approx 1+\frac{2\pi}{\hat{\beta}}$, and 
    \begin{equation}
        \ell = \mathcal{O}\left( \frac{\hat{\beta}}{\epsilon}\right).
    \end{equation}
Therefore, the QSVT approach is more efficient in the finite temperature regime. 
\end{remark}

\bigskip

Polynomial approximations of the density-matrix in \cref{mat-gam1} are not new. In fact, it has been used in \cite{cytter2018stochastic}. But in this classical algorithm, the matrix multiplications will introduce significant computational overhead. In contrast, the quantum singular value transformation (QSVT) \cite{GSLW19}  can efficiently prepare the density-matrix with a complexity that does not depend on the matrix dimension explicitly.   

\begin{lemma}[{\cite[Lemma 48]{GSLW19}}]
  \label{lemma:sparse-to-be}
  Let $H\in\mathbb{C}^{2^m\times 2^m}$ be an $m$-qubit operator with at most $s$ nonzero entries in each row and column. Suppose $H$ is specified by the following sparse-access oracles ${O}_S$ and ${O}_H$ defined in \cref{eq:os,eq:oh}.
  Suppose $\abs{H_{i,j}} \leq 1$ for $i \in [m]$ and $j\in[m]$. Then for all $\epsilon \in (0, 1)$, an $(s, m+3, \epsilon)$-block-encoding of $H$ can be implemented using $\mathcal{O}(1)$ queries to $O_H$ and $O_S$, along with $\mathcal{O}(m+\polylog(1/\epsilon))$ 1- and 2-qubit gates.  
\end{lemma}

\begin{remark}
    According to \cref{sec: relation between max and 2 norms}, the condition that $\abs{H_{i,j}} \leq 1$ for $i \in [m]$ and $j\in[m]$ is automatically satisfied due to the scaling in \cref{eq: scaled quantities}.
\end{remark}

The QSVT builds a block-encoding of the following matrix function,
\begin{equation}\label{be-p}
    U_{p_\ell(H)} = \begin{pmatrix}
        p_{\ell}({H}) & \vdot \\
        \vdot & \vdot\\
    \end{pmatrix},\quad p_{\ell}(x) \approx \left[1+\exp(\hat{\beta}x)\right]^{-1}\text{ on }
    [-1,1],
\end{equation}where $\hat{\beta}$ is defined in \cref{eq: scaled quantities}.  This is summarized as follows,
\begin{lemma}\cite[Theorem 56]{GSLW19}
    Let $U_H$ be a block encoding of $H$. Then there is a quantum circuit $U_{p_\ell(H)}$ 
    which is a block encoding of $f(H)$. The circuit involves $\ell$ application of $U_H$ and $U_H^\dag,$ one application of controlled-$U_H$ gate, and $\mathcal{O}(\ell)$  other one- and two-qubit gates.
\end{lemma}

In light of \cref{lem:ell}, at finite temperature, the complexity of the block encoding only has a logarithmic dependence on $\epsilon.$

\subsection{Estimating the electron density}

Recall that the electron density at different locations corresponds to the diagonals of $f(H)$:
\begin{equation}\label{eq: dens}
    {F}(\bm n)(j) = \tr\big( {\rho}_j f(H) \big).
    \end{equation}
where,
\begin{equation}\label{rho_a}
    \rho_j=  \ketbra{\bm r_j}, \quad j \in [N_I].
\end{equation}
with $\bm r_j$ being an interpolation point in $D_\Delta$. 

The QSVT uses the polynomial approximation ${f(H)} \approx  {p_\ell(H)}$,  and it provides an approximate block encoding of $p_\ell(H)$. Therefore, we use the following estimator for the electron density,
\begin{equation}\label{eq: electron density estimator}
    \hat{F}(\bm n)(j) = \tr\big( {\rho}_j p_\ell(H) \big).
    \end{equation}

In \cref{eq: electron density estimator}, we have treated $p_\ell(H)$ as observables. 
To estimate the expectation in \cref{eq: electron density estimator}, we 
 consider the techniques in Rall \cite{rall2020quantum}. Rall's approach involves the purification of the density operators, the block encodings of the observables, and amplitude amplification \cite{brassard2000amplitude}. Fortunately, the density operators in \cref{rho_a} are pure states, and the observable $f(H)$ is already block-encoded.

\begin{lemma}[\cite{rall2020quantum}]\label{lem:aa}
    If a Hermitian matrix $A$ with $\|A\|_2\leq\alpha$ can be block-encoded by $Q$ elementary gates and a density operator $\rho$ can be purified as 
    \begin{equation}
        \rho=\mathrm{tr}_{\mathbb{C}^{k}}(|\rho\rangle\langle\rho|),\quad |\rho\rangle = U|\bs{0}\rangle|0\rangle_{k}    
    \end{equation}with an unitary $U$ implementable by $R$ elementary gates, then for every $\epsilon,\delta>0$ there exists an algorithm that produces an estimate $\xi$ of $\mathrm{tr}(\rho A)$ such that
    \begin{equation}
        |\xi- \mathrm{tr}(\rho A)|\leq \epsilon,
    \end{equation}with probability $1-\delta$. The gate complexity of the algorithm is $\mathcal{O}\left((R+Q)\frac{\alpha}{\epsilon}\log\frac{1}{\delta}\right)$.
\end{lemma} 

In light of \cref{lem:ell,lem:aa}, we immediately have,

\begin{theorem}\label{prop:Fa}
     For each $j \in [N_I],$ there is a quantum algorithm that outputs an estimate $ \hat{F}(\bm n)(j)$ of $ F(\bm n)(j)$,  with $\epsilon $ accuracy, i.e., \[\abs{\hat{F}(\bm n)(j) - {F}(\bm n)(j)}< \epsilon,\] with probability $1-\delta$. The  algorithm uses  $\mathcal{O}\left(\frac{s }{\epsilon} \log \frac{s}{\epsilon} \log\frac{1}{\delta} \right) $
     queries to ${O}_H.$
\end{theorem}

\subsection{Hybrid algorithms and overall complexity}\label{hybrid}

By far, we have built a procedure for estimating the electron density using QSVT and amplitude amplification on quantum computers. To perform the self-consistent calculation of the DFT, we will use the estimate of the electron density to interface with fixed-point iteration methods on classical computers. Overall, this constitutes a hybrid algorithm for implementing the SCF iteration in the DFT.  An iteration on a classical computer produces a new electron density at the interpolation points in $D_\Delta$. One then evaluates $V_H$ and $V_\text{xc}$ and then interpolates them onto the fine grid in $D_\delta$, as illustrated in \cref{fig:my_label}.  We make the following assumption,

\begin{assumption}
Given the electron density at $N_I$ interpolation points,  the potential $V$ in the Hamiltonian matrix (esp. $V_H$ and $V_\mathrm{xc}$)  can be  evaluated with precision $\epsilon$ 
     with cost $\mathcal{O}(N_I),$ 
     excluding logarithmic factors. 
\end{assumption}

Let us elaborate on this assumption. First, a simple implementation of the interpolation procedure is the multi-grid approach, which has been used in   \cite{merrick1995multigrid} to accelerate the DFT calculations. In this case,  the interpolation points correspond to a  coarse grid. Second, 
the calculation of the exchange-correlation potentials \cite{perdew1981local,perdew1996generalized} at the interpolation points is quite straightforward. Third, the Poisson equation that leads to the Hartree potential can be solved with classical algorithms, e.g., via Fast Fourier transform, which has complexity $\mathcal{O} (N_I\log N_I)$ \cite{braverman1998fast}. It is also possible to solve  Poisson's equation with quantum algorithms \cite{childs2021high,lin2020optimal,cui2023quantum}, in which case the complexity is $\mathcal{O}(sN_I^{2/3}).$ Finally, as we will show in the next section, even without the interpolation step, i.e., $N_I=N_g$, our algorithm still has a cubic speedup over classical algorithms in terms of the number of electrons. Therefore, the computational gain from the interpolation is only moderate, and it is meant for a further reduction of the complexity.

Notice that since the major computational cost in classical algorithms comes from the computation of roughly $N_e$ eigenvalues and eigenvectors,  such an interpolation procedure will not significantly improve the complexity there. In contrast, in the quantum algorithm, the complexity can be mostly attributed to the computation of the expectations, in which case the interpolation provides an important means to reduce the complexity.

\medskip

To quantify convergence, we make a stability assumption.
\begin{assumption}\label{assumption2}
    The Jacobian $\frac{\partial F}{\partial \bm n}(\bm n_*)$ has eigenvalues with real parts less than 1.
\end{assumption}

In this section, we consider two fixed-point methods and show the runtime analysis by establishing the convergence theorems of those methods. The first method is known as the standard fixed-point iteration with simple mixing, which we will call the \emph{ full coordinate fixed-point method} (FCFP), in the sense that the method updates all components of $\hat{F}(\bm n)$ in \cref{eq: dens}. We will show that the iterations converge linearly under suitable conditions. However, the cost for estimating all components of the electron density scales linearly with respect to  $N_I$. As an alternative, we propose a method that requires only some components of $\hat{F}(\bm n)$ to be updated at each iteration. We will call this method the \emph{randomized block coordinate fixed-point method} (RBCFP), which will be made more precise later.

\medskip 

 The convergence of fixed-point iterations usually requires a contraction property of the fixed-point function. For generality, this contraction property is expressed in terms of a weighted vector norm,
 \begin{equation}\label{eq: weighted norm}
    \norm{\bm x}_{\bm w} = \sqrt{\sum_j\bm w(j) \abs{\bm x(j)}^2 }, 
 \end{equation}
where $\bm w\ne \bm 0$ is a vector with nonnegative entries that will be regarded as weights. 

\begin{definition}\label{def: locally-contractive}We say that a mapping $F(\bm n)$ is locally-contractive if there exist some weighted vector norm $\norm{\cdot}_{\bm w}$ and some $\bm w$-dependent $c\in(0,1)$ such that
\begin{equation}\label{eq: contractive}
\begin{aligned}
    &\|F(\bm n)-F(\bm n')\|_{\bm w}\leq c\|\bm n-\bm n'\|_{\bm w},
\end{aligned}
\end{equation}for all $\bm n,\bm n'\in$ $B_{\gamma}(\bm n_*)$ which denotes the ball centered at the fixed point $\bm n_*$ with radius $\gamma$. Here the norm $\|\cdot\|_{\bm w}$ does not have to be the standard Euclidean norm.
\end{definition}

 For the DFT calculations, such property is connected to the structural stability of the underlying physical system \cite{lin2013elliptic,cances2021convergence}. Here we give a mathematical condition based on \cref{assumption2} that ensures a contraction. 
\begin{lemma}[Theorem 3.3 \cite{ko2023stochastic}]\label{lemma: make contraction}
    Under \cref{assumption2}, then there exists a $\gamma>0$, a damping parameter $a \in [0,1]$, an $(a,\frac{\partial F}{\partial\bm n}(\bm n_*))$-dependent weighted norm $\norm{\cdot}_{\bm w}$ and a $\bm w$-dependent $c\in(0,1)$ such that the mapping $ \widetilde{F}  = (1-a)\bm n + aF(\bm n)$ is contractive in the neighboring $B_\gamma(\bm n_*),$ that is,
    \begin{equation}
        \norm{ \widetilde{F}(\bm n_1) - \widetilde{F}(\bm n_2) }_{\bm w} < c  \norm{ \bm n_1 -\bm n_2 }_{\bm w}, \quad 
        \forall \bm n_1,\bm n_2 \in B_\gamma(\bm n_*).
    \end{equation}
\end{lemma}
In fact, the weighted norm $\|\cdot\|_{\bm w}$ is induced by an inner product \cite[Theorem 3.3]{ko2023stochastic}. This fact implies that the weighted norm can be used in place of the standard Euclidean norm in convergence analysis, as shown in \cref{sec: Appendix}, due to the equivalence property of norms in a finite-dimensional Banach space \cite{bressan2012lecture}. In other words, as long as the contraction holds for one vector norm, the convergence property is guaranteed in any other norm.

\subsubsection{The full coordinate fixed-point method}

In this section, we establish the convergence rate of the FCFP method in conjunction with the simple mixing scheme~\cite{lin2013elliptic,cances2021convergence}. \cref{alg: full coordinate fixed-point method} outlined the implementation of the FCFP method. In addition, we present the overall query complexity of the hybrid algorithm equipped with the FCFP method. 

Recall that we denote $\hat{F}(\bm n)$ as the vector in $\mathbb{R}^{N_I}$, whose component is defined by \cref{eq: dens}.
\begin{theorem}\label{thm: full estimation}
    Assume that there exists $a>0$ and $c\in(0,1)$ under the assumption in \cref{lemma: make contraction}. For a given initial guess $\bm n_0\in B_{\gamma}(\bm n_*)$, the FCFP iteration obtained from the simple mixing scheme, 
    \begin{equation}\label{fcfp mapping}
       \bm n_{k+1}=(1-a) \bm n_k+a\hat{F}(\bm n_k),
    \end{equation}
    converges to the fixed-point linearly with probability at least $1-\frac{\|\bm n_0- \bm n_*\|_{\bm w}^2}{\gamma^2}$, 
    \begin{equation}
        \mathbb{E}[\|\bm n_k -\bm  n_*\|_{\bm w}^2] \leq c^{2(k-1)}\|\bm n_0 - \bm n_*\|_{\bm w}^2. 
    \end{equation}
\end{theorem}

The proof of \cref{thm: full estimation} can be found in \cref{sec: proof for full estimation}. Similar results regarding linear convergence have been obtained in \cite{toth_local_2017,lin2013elliptic,cances2021convergence}.

\begin{theorem}\label{thm: full}
    The hybrid algorithm (\cref{alg: full coordinate fixed-point method}) can be implemented  to obtain $\|\bm n_k -\bm  n_*\|_{\bm w} < \epsilon$ with probability at least $1-\delta-\frac{\|\bm  n_0- \bm  n_*\|_{\bm w}^2}{\gamma^2}$   
    with
     \begin{equation}
         \mathcal{O}\left( \frac{sN_I}{\epsilon}  \log \frac{1}{\epsilon}  \log \frac{1}{\delta}\right), 
     \end{equation}
     queries to ${O}_H.$
\end{theorem}

\begin{algorithm}
\SetAlgoLined
	\KwData{initial guess $\bm  n_0$, damping parameter $a\in(0,1)$}
	\KwResult{$\bm n_*$ }

	\For{$k=0:T$}{
             Estimate $\hat{F}(\bm n_k)$ using the QSVT and the amplitude amplification
              
		     $\bm n_{k+1}=(1-a)\bm n_k+a\hat{F}(\bm n_k)$\;
       
       Update the Hamiltonian\;}
	\caption{Full coordinate fixed-point iteration}
    \label{alg: full coordinate fixed-point method}
\end{algorithm}

\subsubsection{The randomized coordinate fixed-point method}

In this section, we introduce an alternative to the FCFP method. Rather than updating all components of $F(\bm n)$,  we only update the components selectively.  The key idea is similar to the randomized coordinate iterative algorithms~\cite{nesterov2012efficiency,tsitsiklis1994asynchronous,peng2016arock}. The new method will be termed the randomized coordinate fixed-point method (RCFP). The basic steps are outlined in \cref{alg: randomized coordinate fixed-point method}. Formally, we define the RCFP method

\begin{definition}
       Given a fixed-point mapping $\hat{F}(\bm n) $, a randomized block coordinate fixed-point mapping (RBCFP) is defined as
    \begin{equation}\label{eq: randomized coordinate fixed-point mapping 2}
        \hat{F}_{R,m}(\bm n) = 
        \sum_{k\in \{k_{R_j}\}_{j=1}^m}(\bs{u}_k, \hat{F}(\bm n))\bs{u}_k
        + \sum_{k\not\in \{k_{R_j}\}_{j=1}^m}(\bs{u}_k,\bm n)\bs{u}_k,
    \end{equation}where $\{k_{R_j}\}_{j=1}^m$ is the set of m indices randomly sampled from the index set $[N_I]$, uniformly without replacement, 
    and the parenthesis $(\;,\;)$ refers to the standard inner product between vectors. 
\end{definition}We remark that the method in \cref{thm: full estimation} corresponds to the special case $m=N_I$. The following theorem shows that despite the partial update of the density, the method still has linear convergence.

\begin{theorem}\label{thm: randomized estimation 2}
 Assume that there exist $a>0$ and $c\in(0,1)$ as in \cref{lemma: make contraction}. Let $m\in\{2,..,N_I\}$ be given. For a given initial guess $n_0\in B_{\gamma}(n_*)$, the RBCFP iteration obtained from the simple mixing scheme,
 \begin{equation}
   \bm n_{k+1}=(1-a)\bm n_k+a\hat{F}_{R,m}(\bm n_k)
 \end{equation}
 converges to the fixed-point linearly with probability at least $1-\frac{\|\bm n_0-\bm n_*\|_{\bm w}^2}{\gamma^2}$,
    \begin{equation}
         \mathbb{E}[\|\bm n_k-\bm n_*\|_{\bm w}^2]\leq \left(1-\frac{m(1-c^2)}{N_I}\right)^k\|\bm n_0-\bm n_*\|_{\bm w}^2.
    \end{equation}
\end{theorem}
The proof of \cref{thm: randomized estimation 2} can be found in \cref{sec: proof for randomized estimation 2}. 

\begin{remark}
It is worthwhile to highlight the differences between the FCFP and RBCFP methods. First, the admissible range of the damping parameter in \cref{thm: randomized estimation 2} can be different from that of the damping parameter in \cref{thm: full estimation}. This is because the Jacobian of the mapping in the RBCFP method (\cref{eq: randomized coordinate fixed-point mapping 2}) is different from that of the FCFP method (\cref{fcfp mapping}). More precisely, the RBCFP method can perform with a larger damping parameter without the loss of stability. Second, the convergence rate in \cref{thm: randomized estimation 2} is proven for the worst-case scenario with the same choice of the damping parameter in \cref{thm: full estimation}. In practice, we expect that the RBCFP has the potential for faster convergence. This has been observed in our numerical results in \cref{sec:num}.  
\end{remark}

\begin{theorem}\label{thm: rand}
    The hybrid algorithm (\cref{alg: randomized coordinate fixed-point method}) can be implemented  $\|\bm n_k -\bm n_*\|_{\bm w} < \epsilon$ with probability at least $1-\delta -\frac{\|\bm n_0-\bm n_*\|_{\bm w}^2}{\gamma^2}$  with 
     \begin{equation}
         \mathcal{O}\left( \frac{s N_I}{\epsilon} \log\frac{1}{\epsilon} \log \frac{1}{\delta}\right), 
     \end{equation}
     queries to ${O}_H.$
\end{theorem}

\begin{algorithm}
\SetAlgoLined
	\KwData{initial guess $n_0$, damping parameter $a\in(0,1)$,  index parameter $m\in[N_I]$}
	\KwResult{$n_*$ }

	\For{$k=0:T$}{
             Sample $m$ different indices $\{k_{R,j}\}_{j=1}^m\subset [N_I]$ uniformly
 
             Estimate $\hat{F}_{R,m}[n_k]$ using the QSVT and the Amplitude Amplification.
              
		     $n_{k+1}=(1-a)n_k+a\hat{F}_{R,m}[n_k]$\;
       
        Update the Hamiltonian\;}
	\caption{Randomized block coordinate fixed-point iteration}
    \label{alg: randomized coordinate fixed-point method}
\end{algorithm}

\subsubsection{Estimating the Chemical potential}

Within the hybrid algorithms in \cref{alg: full coordinate fixed-point method,alg: randomized coordinate fixed-point method}, we have so far focused on the case with given chemical potential $\mu$. If $N_e$ is given instead, we can incorporate the constraint in \cref{constraint} to determine $\mu$. At the continuous level, this implies that
\begin{equation}
    \int n(\bs{r})\, \dd\bs{r} = N_e,\quad n(\bs{r}) = \langle\bs{r}|f(H-\mu I)|\bs{r}\rangle. 
\end{equation}
The first equation can be cast into a nonlinear equation, 
\begin{equation}\label{eq: continuous G}
    G(n, \mu)=0, \quad G(n, \mu):= \int n(\bs{r}) \,\dd\bs{r} - N_e.
\end{equation}
Given $n(\bm r)$, $G$ is a monotone function of $\mu.$

To incorporate the constraint in \cref{constraint} in our quantum algorithm, we update the chemical potential on classical computers together with the update of the electron density, e.g., in \cref{alg: randomized coordinate fixed-point method}. This extended algorithm consists of the following steps,
\begin{equation}\label{eq: update n and mu}
    \begin{split}
       \bm  n_{k+1} &= (1-a) \bm  n_k+a\hat{F}_{R,m}(\bm  n_k),\\
        \mu_{k+1} &= \mu_k - \eta \hat{G}(\bm n_{k+1}), \quad 
    \end{split}
\end{equation}
where 
\[
        \hat{G}(\bm n) \coloneqq \sum_j  \abs{\bm n(j)}   \delta^3 - N_e,
\]
is a discretization of \cref{eq: continuous G}.

Here $\delta^3$ is the infinitesimal volume from the finite-difference approach with grid size $\delta$; $\eta \in(0,1)$ is the damping parameter for updating the Fermi energy. 
This solver for $\mu$ is motivated by the stochastic approximation method by Robbins and Monro for solving nonlinear equations \cite{robbins1951stochastic}.

\section{Numerical Results}\label{sec:num}

\subsection{Experiment details}
To mimic our hybrid quantum algorithm on a classical computer, we conducted numerical tests for the approximation of the density-matrix in \cref{eq: electron density estimator} within the MATLAB platform M-SPARC, a real-space density functional electronic structure code~\cite{ghosh2017sparc}. We chose Barium titanate (BaTiO3) and a water molecule H$_2$O-sheet as our test models from the set of examples in M-SPARC \footnote[1]{https://github.com/SPARC-X/M-SPARC/tree/master/tests}. In the models, temperatures are set to $T=300K$ for BaTiO3 and $T=2320$ for H$_2$O, respectively. The BaTiO3 system is set up in a supercell in a cubic domain with periodic boundary conditions. The H$_2$O system is treated with periodic boundary conditions in the $x-y$ plane where the three atoms are positioned and a Dirichlet boundary condition in the $z$ direction. The local density approximation (LDA) is used for exchange and correlation. We should point out that the M-SPARC code uses a pseudopotential, which we did not consider in our quantum algorithm. Our emphasis, however, is to use the corresponding Hamiltonian $H$ to test the polynomial approximation of the density-matrix, and more importantly, the convergence of the SCF iterations. 

The initial electron density $\bm n_{0}$ in M-SPARC is given as a sum of isolated atom densities.
We perform the calculation of the ground state electron density with either a given chemical potential $\mu$ or by fixing a number of electrons $N_e$. In all tests, the ground truth, i.e., $\bm n_*$ is the converged electron density obtained from the simple mixing scheme of SCF iteration, based on the Fermi-Dirac smearing and direct eigenvalue computation in M-SPARC. In monitoring the convergence of the SCF iterations, we measure the error between the true density $\bm n_*$ and one obtained from FCFP or RBCFP, i.e., $\bm n_k$ together with the Chebyshev approximation method (see \cref{alg: full coordinate fixed-point method} and \cref{alg: randomized coordinate fixed-point method}).

 \subsection{The efficiency of the RBCFP method}
\begin{figure}[thbp]
    \centering
    \begin{subfigure}[b]{0.4\textwidth}
    \centering
    \includegraphics[width=\textwidth]{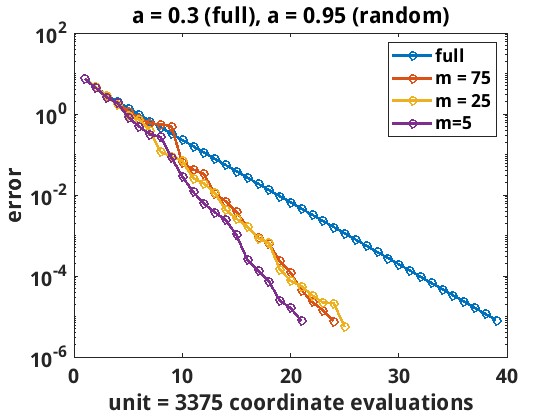}
    \end{subfigure}
   \begin{subfigure}[b]{0.4\textwidth}
    \centering
    \includegraphics[width=\textwidth]{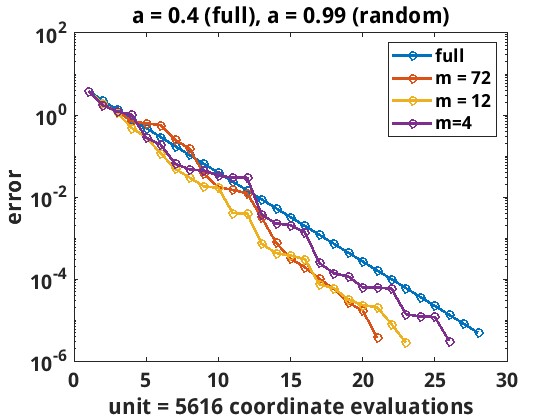}
    \end{subfigure}
    \caption{Comparison of the FCFP and RBCFP methods from \cref{alg: full coordinate fixed-point method,alg: randomized coordinate fixed-point method}. In both panels, the $x$-axis labels the number of coordinate evaluations. The $y$-axis labels the error of the electron density on a logarithmic scale.
    Left: BaTiO3 system with $N_e=$ 40 electrons fixed; Right H$_2$O with $N_e=8$ electrons fixed.    In the left panel, the FCFP method runs with damping parameter $a=0.3$, but the RBCFP method with $a=0.95$. In the right panel, the FCFP runs with $a=0.4$, but the RBCFP with $a=0.99$. }
    \label{fig: experiment}
\end{figure}

 To first fully focus on the performance of the FCFP and RBCFP methods, we computed the density matrix in \cref{mat-gam1} exactly as shown in \cref{fig: experiment}.  For each of the two physical systems, we run the RBCFP with three different block sizes and then compare the convergence to that of the FCFP method. The error is shown on a logarithmic scale in the figure.  To compare the performance on an equal footing, we rescaled the $x$ axis to indicate the number of coordinate evaluations. The SCF iterations were terminated when the relative error between the electron density and the true one is below $10^{-6}$ as default in M-SPARC. There are several interesting aspects to note from the results in \cref{fig: experiment}. First, while it is well-known that the simple mixing scheme of the direct SCF calculations leads to linear convergence \cite{lin2013elliptic,cances2021convergence}, the RBCFP method also exhibits linear convergence, which supports our theoretical results \cref{thm: full estimation} and \cref{thm: randomized estimation 2}. Second, for the convergence of both methods, it is important to select proper damping parameters. For example, in \cref{tab: damping selection}, we checked different damping parameters for the two systems and found the best damping parameters for the FCFP method in terms of the number of iterations until convergence, where the optimal values are found to be around 0.4 for both test cases. However, as shown in \cref{fig: experiment}, it turns out that the RBCFP method can perform well with much larger damping parameters that are very close to 1. A similar observation has been made in the context of coordinate descent optimization methods in machine learning~\cite{nutini2015coordinate,nesterov2012efficiency}. 
 In addition, \cref{fig: experiment} shows that the RBCFP method can converge faster than the FCFP method by an order of 2 (BaTiO3) and 1.5 (H$_2$O), which supports the different convergence rates proven in \cref{thm: full estimation,thm: randomized estimation 2}. From the efficiency of the RBCFP method shown in \cref{fig: experiment}, we highlight that the practice of updating only a few coordinates randomly selected at each iteration step can result in the estimation of only a few diagonal elements from quantum computation in our hybrid algorithm, which amounts to a reduction of the overall complexity.

\begin{table}[htbp]
    \centering
    \begin{subtable}[h]{0.45\textwidth}
        \centering
         \begin{tabular}{c|c|c|c|c|c}
        Damping parameter & {\color{red}0.3} & 0.33 & 0.35 & 0.37 & 0.38\\
        \hline
        SCF iterations & 39 & 40 & 76 & 482 & diverge \\
    \end{tabular}
    \caption{system BaTiO3}
    \end{subtable}
   \begin{subtable}[h]{0.45\textwidth}
        \centering
         \begin{tabular}{c|c|c|c|c|c}
        Damping parameter & 0.3 & {\color{red}0.4} & 0.51 & 0.55 & 0.58 \\
        \hline
        SCF iterations & 39 & 28 & 30 & 112 & diverge
    \end{tabular}
    \caption{system H$_2$O}
    \end{subtable}
    \caption{The role of the damping parameter in the convergence of direct SCF iterations. The table shows the number of SCF iterations for relative error $10^{-6}$ with the simple mixing scheme applied to the exact SCF formulation in \cref{fixed-point-problem3} for the given damping parameters for two systems BaTiO3 (Left) and H$_2$O (Right). }
    \label{tab: damping selection}
\end{table}

\begin{figure}[htbp]
   \centering
   \begin{subfigure}[b]{0.54\textwidth}
    \centering
    \includegraphics[width=\textwidth]{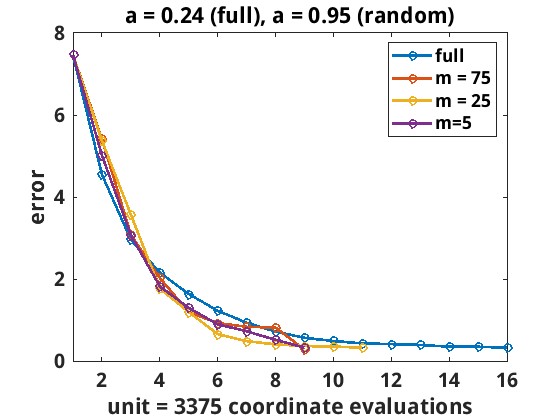}
    \end{subfigure}
    \caption{Comparison of the FCFP and RBCFP methods with block sizes $m=5,25,75$ using the Chebyshev approximation method in \cref{eq: electron density estimator} for system BaTiO3. The degree of the method is 500. The $x$-axis labels the number of coordinate evaluations. The $y$-axis is the error of the electron density. The FCFP method runs with damping parameter $a=0.24$, but the RBCFP method with $a=0.95$.}
    \label{fig: experiment 2}
\end{figure}

\begin{figure}[htbp]
   \centering
    \begin{subfigure}[b]{0.4\textwidth}
    \centering
    \includegraphics[width=\textwidth]{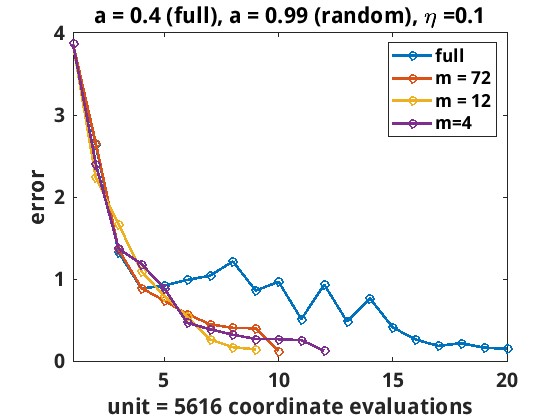}
    \end{subfigure}
    \begin{subfigure}[b]{0.4\textwidth}
    \centering
    \includegraphics[width=\textwidth]{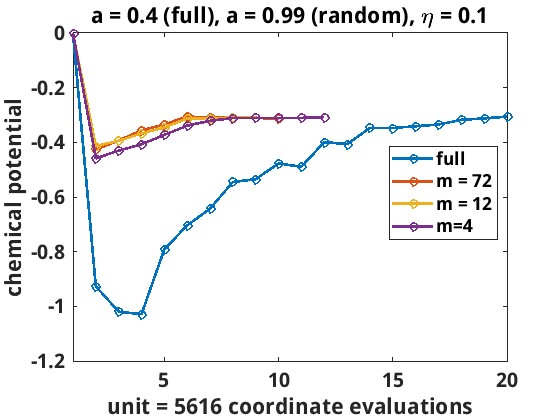}
    \end{subfigure}
    \caption{Comparison of the FCFP and RBCFP methods with block sizes $m=4,12,72$ using the Chebyshev approximation method in \cref{eq: electron density estimator} for system H$_2$O-sheet. The degree of the method is 500. The $x$-axis labels the number of coordinate evaluations. Left: the $y$-axis is the error of the electron density. Right: the values of the chemical potential $\mu$ during the iterations. The FCFP method runs with damping parameter $a=0.4$, but the RBCFP method with $a=0.99$. The damping parameter for the chemical potential in \cref{eq: update n and mu} is $\eta = 0.1$. }
    \label{fig: experiment 3}
\end{figure}

Our next numerical experiment incorporates the Chebyshev polynomial approximation of the density-matrix, which mimics the QSVT implementation of the density-matrix on a classical computer. For the system BaTiO3, we applied the Chebyshev approximation method with degree $\ell=500$ as \cref{eq: electron density estimator} for implementing the RBCFP method. We used a fixed chemical potential $\mu = 0.3403 $(eV) that is associated with the ground truth $\bm n_*$ used in \cref{fig: experiment}. In \cref{fig: experiment 2}, we observe that the RBCFP methods still converge faster than the FCFP method in terms of coordinate evaluations to a given precision. Similarly in \cref{fig: experiment 3}, we applied the polynomial approximation method for system H$_2$O-sheet. One difference is that we used the variable chemical potential in \cref{eq: update n and mu} to satisfy the constraint on $N_e$ during the iteration. Still, we can clearly see that the RBCFP methods converge to a given precision faster than the FCFP method in terms of the electron density. Furthermore, it is observed that the chemical potentials obtained from the RBCFP method converge faster than one from the FCFP method.

One interesting observation in \cref{fig: experiment 2} was that when we used the same damping parameter for the FCFP method as in \cref{fig: experiment}, it could not reach the given precision. We numerically found $a=0.24$ as the nearly optimal value to reach the precision. However, the RBCFP implementations with the same damping parameter still converge well. This might be attributed to the fact that the Jacobian at $\bm n_*$ is defined by the polynomial matrix function, rather than the Fermi-Dirac function, and the upper bound of damping parameters for the FCFP is altered. For more rigorous results, we leave this observation to future work.

\section{Summary and Discussions}

We proposed an algorithm for the density-functional theory with complexity that scales linearly with the dimension of the density update $F(\bm n)$  which is often much less than the number of electrons. Therefore, this can be considered as linear/sublinear scaling, which compared to the cubic scaling in classical algorithms, is a significant reduction. 

The first natural question is whether the current algorithm can be improved to have a better dependence on the dimension of the density update  $F(\bm n)$.  There are quantum algorithms that offer quadratic speedup, e.g., the gradient estimation approach by Huggins et al.~\cite{HMW+21} for estimating multiple observables. But our formulation in  \cref{eq: electron density estimator} is based on a single observable with multiple density operators. It is not yet clear whether these algorithms can be applied. 

As pointed out in \cref{low-temp}, the degree of the polynomial in the approximation of the Fermi-Dirac function increases considerably for lower temperature values. This is due to the fact that in this regime the  Fermi-Dirac function approaches a step function, which is discontinuous. In this regime, $\ell$ must be proportional to $\hat{\beta}/\epsilon$, and the overall complexity increases significantly. 

Another common practice in DFT calculations is to exclude core electrons and incorporate their effects by using pseudopotentials. Although it is not clear whether this practice is needed in a quantum algorithm, it is still of theoretical interest to explore how such potentials can be block encoded into $U_H$. These issues will be explored in separate works. 

\section*{Acknowledgement}
XL's research is supported by the National Science Foundation Grants DMS-2111221. CW acknowledges support from National Science Foundation grant CCF-2238766 (CAREER). Both XL and CW were supported by a seed grant from the Institute of Computational and Data Science (ICDS) and the National Science Foundation Grants CCF-2312456. 

\crefalias{section}{appsec}
\section{Appendices}
\label[appsec]{sec: Appendix}

\subsection{Relation between the max norm and 2-norm of square matrix}\label[appsec]{sec: relation between max and 2 norms}

We show that for any square matrix $A\in\mathbb{C}^{N\times N}$, 
\begin{equation}
    \|A\|_{\max} \leq \|A\|_2,
\end{equation}where $\|A\|_{\max}:=\max_{i,j}|A_{ij}|$ and $\|A\|_2$ is the 2-norm of $A$. Let $A=U\Sigma V^{\dagger}$ be the singular value decomposition of $A$. Denote $x_i=U^{\dagger}e_i$ and $y_i=V^{\dagger}e_i$ for each $i\in [N]$. By the Cauchy-Schwarz inequality, we observe that
\begin{equation}
    \begin{split}
        |A_{ij}|^2 &= |e_i^TAe_j|^2 = |x_i^{\dagger}\Sigma y_j|^2=|\sum_{k=1}^N\sigma_k(\bar{x}_i)_k(y_j)_k|^2\\
        &\leq \left(\sum_k\sigma_k|(x_i)_k|^2\right)\left(\sum_k\sigma_k|(y_j)_k|^2\right)\leq \|A\|_2^2\|x_i\|_2^2\|y_j\|_2^2 = \|A\|_2^2,
    \end{split}
\end{equation}which proves the statement.

\subsection{Proof of \cref{thm: full estimation}}\label[appsec]{sec: proof for full estimation}

We define the fixed-point mapping as follows 
\begin{equation}\label{eq: estimate of all components of n}
    \hat{F}(\bm n_{t+1}) = \sum_{j=1}^{N_I}\langle \bs{r}_{j}|P_{\xi_{t,j}}p_{\ell}(H_t)P_{\xi_{t,j}}|\bs{r}_{j}\rangle\bs{u}_{j}.
\end{equation}At iteration $t$, $\bs{u}_{j}$ is the $j$-th standard basis vector, $P_{\xi_{t,j}}$ stands for the measurement projector of $H_t$ corresponding to index $j$ and $\bs{r}_{j}$ is the grid point. In other words, the quantity in \cref{eq: estimate of all components of n} is an unbiased estimate for the update of the electron density.

Now we consider the iteration defined as,
\begin{equation}\label{eq: fixed-point iteration full coordinate}
   \bm n_{t+1} = (1-a)\bm n_t+a\hat{F}(\bm n_t).
\end{equation}
Define the characteristic function $\mathbb{I}_t$ that values $1$ if the first $p$-th iterates stay in $B_{\gamma}(\bm n_*)$ and otherwise $0$. We denote by $\mathbb{E}_t:=\mathbb{E}[\cdot||\mathbb{I}_t=1]$, the expectation conditioned on an event that $\mathbb{I}_t=1$. Let $\bs{e}_t=\bm n_t-\bm n_*$ be the error between the current iterate and the fixed point.

The fixed-point iteration in \cref{eq: fixed-point iteration full coordinate} yields a recursive inequality as follows,
\begin{equation}\label{eq: recursive ineq}
\begin{split}
    \mathbb{E}_t[\|\bs{e}_{t+1}\|_{\bm w}^2\mathbb{I}_{t+1}]&\leq \mathbb{E}_t[\|\bs{e}_{t+1}\|_{\bm w}^2\mathbb{I}_{t}]\\
    &\leq (1-a)^2\|\bs{e}_t\|_{\bm w}^2+2a(1-a)\mathbb{E}_t\left[\left(\bs{e}_t,\hat{F}(\bm n_t)-\bm n_*\right)_{\bm w}\right]+a^2\mathbb{E}_t[\|\hat{F}(\bm n_t)-\bm n_*\|_{\bm w}^2]
\end{split}
\end{equation}
The middle term of the right-hand side in \cref{eq: recursive ineq} is simplified as
\begin{equation}
    \begin{split}
        &\mathbb{E}_t\left[\left(\bs{e}_t,\hat{F}(\bm n_t)-\bm n_*\right)_{\bm w}\right] = \left(\bs{e}_t,\sum_{j=1}^{N_I}\langle \bs{r}_j|p_{\ell}(H_t)|\bs{r}_j\rangle\bs{u}_j-\bm n_*\right)_{\bm w}.
    \end{split}
\end{equation}

To simplify this inequality, we first observe that for every $j\in[M]$,
\begin{equation}\label{eq: takeout square}
    \begin{split}
         \mathbb{E}_t[(\bs{u}_j,\hat{F}(\bm n_t))^2]&=
         \mathbb{E}_t\left[\left(\langle \bs{r}_{j}|P_{\xi_{t,j}}p_{\ell}(H_t)P_{\xi_{t,j}}|\bs{r}_{j}\rangle\right)^2\right]\\
         &= \sum_{\xi = 1}^{N_I}(\langle \bs{r}_{j}|P_{\xi}p_{\ell}(H_t)P_{\xi}|\bs{r}_{j}\rangle)^2\cdot \|P_{\xi}|\bs{r}_{j}\rangle\|_2^2\\
        & = \sum_{\xi = 1}^{N_I}p_{\ell}(\lambda_{t,\xi})^2\cdot \|P_{\xi}|\bs{r}_{j}\rangle\|_2^4\\
        &\leq \left(\sum_{\xi = 1}^{N_I}p_{\ell}(\lambda_{t,\xi})\cdot \|P_{\xi}|\bs{r}_{j}\rangle\|_2^2\right)^2 \quad p_{\ell}(x)\approx f(x)>0\\
        & = \left(\mathbb{E}_t[\langle \bs{r}_{j}|P_{\xi_{t,j}}p_{\ell}(H_t)P_{\xi_{t,j}}|\bs{r}_{j}\rangle]\right)^2\\
        & = \left(\mathbb{E}_t\left[(\bs{u}_j,\hat{F}(\bm n_t))\right]\right)^2
    \end{split}
\end{equation}
By this result, the last term of the right hand side in \cref{eq: recursive ineq} can be estimated as
\begin{equation}\label{eq: last term in full coordinate}
\begin{split}
    \mathbb{E}_t[\|\hat{F}(\bm n_t)-\bm n_*\|_{\bm w}^2 &= \sum_{j=1}^{N_I}\bm w(j)\mathbb{E}_t\left[\left(\bs{u}_j,\hat{F}(\bm n_t)-\bm n_*\right)^2\right]\\
    & = \sum_{j=1}^{N_I}\bm w(j)\mathbb{E}_t\left[(\bs{u}_j,\hat{F}(\bm n_t))^2-2(\bs{u}_j,\hat{F}(\bm n_t))(\bs{u}_j,\bm n_*)+(\bs{u}_j,\bm n_*)^2\right]\\
    &\leq \sum_{j=1}^{N_I}\bm w(j)\left[\left(\mathbb{E}_t\left[\bs{u}_j,\hat{F}(\bm n_t)\right]\right)^2-2\mathbb{E}_t[(\bs{u}_j,\hat{F}(\bm n_t))](\bs{u}_j,\bm n_*)+(\bs{u}_j,\bm n_*)^2\right]\\
    & = \sum_{j=1}^{N_I}\bm w(j)\left(\mathbb{E}_t\left[(\bs{u}_j,\hat{F}(\bm n_t))\right]-(\bs{u}_j,\bm n_*)\right)^2\\
    & = \|\sum_{j=1}^{N_I}\langle \bs{r}_j|p_{\ell}(H_t)|\bs{r}_j\rangle\bs{u}_j-\bm n_*\|_{\bm w}^2,
\end{split}
\end{equation}where $\bm w(j)$'s are defined in \cref{eq: weighted norm}. Therefore, we can reduce \cref{eq: recursive ineq} to
\begin{equation}\label{eq: contractive ineq}
    \mathbb{E}_t[\|\bs{e}_{t+1}\|_{\bm w}^2\mathbb{I}_{t+1}]\leq \left\|(1-a)\bs{e}_t+a\left(\sum_{j=1}^{N_I}\langle \bs{r}_j|p_{\ell}(H_t)|\bs{r}_j\rangle\bs{u}_j-\bm n_*\right)\right\|_{\bm w}^2 = \|F(\bm n_t)-\bm n_*\|_{\bm w}^2\leq c^2\|\bs{e}_t\|_{\bm w}^2,
\end{equation}where $c$ is defined in \cref{def: locally-contractive}. This proves \cref{thm: full estimation}.

\subsection{Convergence of the RBCFP when $m=1$}\label[appsec]{sec: proof for randomized estimation}

We recall the RBCFP method in \cref{eq: randomized coordinate fixed-point mapping 2} with $m=1$, namely,  
\begin{equation}\label{eq: estimate of component of n}
   \hat{F}_R(\bm n_t) = (\bs{u}_{k_t},\hat{F}(\bm n_t))\bs{u}_{k_t} + \sum_{k\neq k_t}(\bs{u}_k,\bm n_t)\bs{u}_k,
\end{equation}where $k_t$ denotes the index sampled at iteration $i$. Specifically, the estimated component is expressed as 
\begin{equation}
    (\bs{u}_{k_t},\hat{F}(\bm n_t))=\langle \bs{r}_{k_t}|P_{\xi_t}p_{\ell}(H_t)P_{\xi_t}|\bs{r}_{k_t}\rangle,
\end{equation}which is the $k_t$-th component of the full coordinate estimation in \cref{eq: estimate of all components of n}. Here $\xi_t\in[N_I]$ denotes the index corresponding to measurement. 

Noticing the randomness of the RBCFP method from sampling index, we observe that 
\begin{equation}\label{eq: observation}
    \begin{split}
        \mathbb{E}_t[\hat{F}_R(\bm n_t)] &= \mathbb{E}_{k_t}\left[\mathbb{E}_{\xi_t}\left[(\bs{u}_{k_t},\hat{F}(\bm n_t))\bs{u}_{k_t} + \sum_{k\neq k_t}(\bs{u}_k,\bm n_t)\bs{u}_k\right]\right]\\
        & = \mathbb{E}_{k_t}\left[\langle \bs{r}_{k_t}|p_{\ell}(H_t)|\bs{r}_{k_t}\rangle\bs{u}_{k_t}+\sum_{k\neq k_t}(\bs{u}_k,\bm n_t
        )\bs{u}_k\right]\quad\text{remove quantum noise}\\
        & = \mathbb{E}_{k_t}\left[\langle \bs{r}_{k_t}|p_{\ell}(H_t)|\bs{r}_{k_t}\rangle\bs{u}_{k_t}\right]+\frac{N_I-1}{N_I}\bm n_t.
    \end{split}
\end{equation}

Similar to the mixing scheme in \cref{eq: fixed-point iteration full coordinate}, we consider the following iteration,
\begin{equation}
    \bm n_{t+1} = (1-a)\bm n_t+a\hat{F}_R(\bm n_t).
\end{equation}
From this iteration, we have
\begin{equation}\label{eq: recursive ineq2}
\begin{split}
    \mathbb{E}_t[\|\bs{e}_{t+1}\|_{\bm w}^2\mathbb{I}_{t+1}]&\leq \mathbb{E}_t[\|\bs{e}_{t+1}\|_{\bm w}^2\mathbb{I}_t]\\
    & = (1-a)^2\|\bs{e}_t\|_{\bm w}^2+2a(1-a)\mathbb{E}_t\left[\left(\bs{e}_t,\hat{F}_R(\bm n_t)-\bm n_*\right)_{\bm w}\right]+a^2\mathbb{E}_t\left[\|\hat{F}_R(\bm n_t)-\bm n_*\|_{\bm w}^2\right].
\end{split}
\end{equation}By the observation (\cref{eq: observation}), we first simplify the middle term of the right hand side in \cref{eq: recursive ineq2} as follows
\begin{equation}
\begin{split}
    \mathbb{E}_t\left[\left(\bs{e}_t,\hat{F}_R(\bm n_t)-\bm n_*\right)_{\bm w}\right]&= \left(\bs{e}_t,\mathbb{E}_t\left[\hat{F}_R(\bm n_t)-\bm n_*\right]\right)_{\bm w} \\
    & = \frac{N_I-1}{N_I}\|\bs{e}_t\|_{\bm w}^2+(\bs{e}_t,\mathbb{E}_{k_t}\left[\langle \bs{r}_{k_t}|p_{\ell}(H_t)|\bs{r}_{k_t}\rangle\bs{u}_{k_t}-\bm n_*(k_t)\bm u_{k_t}\right])_{\bm w}\\
    & = \frac{N_I-1}{N_I}\|\bs{e}_t\|_{\bm w}^2+\mathbb{E}_{k_t}\left[\bm w(k_t)\bm e_t(k_t)\left(\langle \bs{r}_{k_t}|p_{\ell}(H_t)|\bs{r}_{k_t}\rangle-\bm n_*(k_t)\right)\right].
\end{split}
\end{equation}
We estimate the last term of the right hand side in \cref{eq: recursive ineq2} as follows,
\begin{equation}\label{eq: last term in randomized coordinate}
    \begin{split}
        &\mathbb{E}_t\left[\|\hat{F}_R(\bm n_t)-\bm n_*\|_{\bm w}^2\right] \\
        &= \mathbb{E}_t\left[\bm w(k_t)\left(\hat{F}(\bm n_t)-\bm n_*,\bs{u}_{k_t}\right)^2+\sum_{k\neq k_t}\bm w(k)(\bs{e}_t,\bs{u}_k)^2\right]\\
        & = \mathbb{E}_t\left[\bm w(k_t)\left(\hat{F}(\bm n_t)-\bm n_*,\bs{u}_{k_t}\right)^2\right]+\frac{N_I-1}{N_I}\|\bs{e}_t\|_{\bm w}^2\\
        & \leq \mathbb{E}_{k_t}\left[\bm w(k_t)\left(\mathbb{E}_{\xi_t}\left[(\bs{u}_{k_t},\hat{F}(\bm n_t))\right]-(\bs{u}_{k_t},\bm n_*)\right)^2\right] +\frac{N_I-1}{N_I}\|\bs{e}_t\|_{\bm w}^2\quad\text{similar to } \cref{eq: last term in full coordinate}\\
        &= \mathbb{E}_{k_t}\left[\bm w(k_t)\left(\langle \bs{r}_{k_t}|p_{\ell}(H_t)|\bs{r}_{k_t}\rangle-(\bs{u}_{k_t},\bm n_*)\right)^2\right] +\frac{N_I-1}{N_I}\|\bs{e}_t\|_{\bm w}^2\quad\text{remove quantum noise}.
    \end{split}
\end{equation}
Note that the first term $\|\bm e_t\|_{\bm w}^2$ is expressed as
\begin{equation}
    \|\bm e_t\|_{\bm w}^2 = \frac{N_I-1}{N_I}\|\bm e_t\|_{\bm w}^2+\mathbb{E}_{k_t}[\bm w(k_t)\bm e(k_t)^2].
\end{equation}
To put together all results, we can simplify \cref{eq: recursive ineq2} as
\begin{equation}\label{eq: contractive ineq2}
\begin{split}
    \mathbb{E}_t[\|\bs{e}_{t+1}\|_{\bm w}^2\mathbb{I}_{t+1}]&\leq \frac{N_I-1}{N_I}\|\bs{e}_t\|_{\bm w}^2 + \mathbb{E}_{k_t}\left[\bm w(k_t)\left[(1-a)\bm e_t(k_t)+a(\langle \bs{r}_{k_t}|p_{\ell}(H_t)|\bs{r}_{k_t}\rangle-\bm n_*(k_t))\right]^2\right]\\
    &\leq \frac{N_I-1}{N_I}\|\bs{e}_t\|_{\bm w}^2 + \frac{1}{N}\left\|(1-a)\bs{e}_t+a\left(\sum_{j=1}^{N_I}\langle \bs{r}_j|p_{\ell}(H_t)|\bs{r}_j\rangle\bs{u}_j-\bm n_*\right)\right\|_{\bm w}^2\\
    &\leq \frac{N_I-1}{N_I}\|\bs{e}_t\|_{\bm w}^2+\frac{c^2}{N_I}\|\bs{e}_t\|_{\bm w}^2\\
    &\leq \left(1-\frac{1-c^2}{N_I}\right)\|\bs{e}_t\|_{\bm w}^2,
\end{split}
\end{equation}which completes the proof of the convergence of the RBCFP with $m=1$.

\subsection{Proof of \cref{thm: randomized estimation 2}}\label[appsec]{sec: proof for randomized estimation 2}

The key idea for proving \cref{thm: randomized estimation 2} is not very different from the proof in \cref{sec: proof for randomized estimation}. The main difference is that the error analysis (\cref{eq: recursive ineq2}) now involves the term $\hat{F}_{R,m}$ for a given $m\in[N_I]$ in the middle and last terms.  

We first observe that
\begin{equation}
    \begin{split}
        \mathbb{E}_t\left[\hat{F}_{R,m}(\bm n_t)\right] &= \mathbb{E}_{\{k_{t,j}\}_{j=1}^m}\left[\mathbb{E}_{\{\xi_{t,j}\}_{j=1}^m}\left[\sum_{j=1}^m(\bs{u}_{k_{t,j}},\hat{F}(\bm n_t))\bs{u}_{k_{t,j}} + \sum_{k\not\in \{k_{t,j}\}_{j=1}^m}(\bs{u}_k,\bm n_t)\bs{u}_k\right]\right]\\
        & = \mathbb{E}_{\{k_{t,j}\}_{j=1}^m}\left[\sum_{j=1}^m\langle \bs{r}_{k_{t,j}}|p_{\ell}(H_t)|\bs{r}_{k_{t,j}}\rangle\bs{u}_{k_{t,j}} + \sum_{k\not\in \{k_{t,j}\}_{j=1}^m}(\bs{u}_k,\bm n_t)\bs{u}_k\right] \\
        & =\mathbb{E}_{\{k_{t,j}\}_{j=1}^m}\left[\sum_{j=1}^m\langle \bs{r}_{k_{t,j}}|p_{\ell}(H_t)|\bs{r}_{k_{t,j}}\rangle\bs{u}_{k_{t,j}}\right] + \frac{{N_I \choose m}-{N_I-1 \choose m-1}}{{N_I \choose m}}\bm n_t\\
        & = \mathbb{E}_{\{k_{t,j}\}_{j=1}^m}\left[\sum_{j=1}^m\langle \bs{r}_{k_{t,j}}|p_{\ell}(H_t)|\bs{r}_{k_{t,j}}\rangle\bs{u}_{k_{t,j}}\right] + \frac{N_I-m}{N_I}\bm n_t,
    \end{split}
\end{equation}where the expectations $\mathbb{E}_{\{k_{t,j}\}_{j=1}^m}$ and $\mathbb{E}_{\{\xi_{t,j}\}_{j=1}^m}$ are performed with respect to the index sampling and quantum noise, respectively. From this result, we achieve a slight modification of the recursive inequality in \cref{eq: contractive ineq2} as follows
\begin{equation}\label{eq: contractive ineq3}
    \begin{split}
        \mathbb{E}_t[\|\bs{e}_{t+1}\|_{\bm w}^2\mathbb{I}_{t+1}]&\leq \frac{N_I-m}{N_I}\|\bs{e}_t\|_{\bm w}^2 + \mathbb{E}_{\{k_{t,j}\}_{j=1}^m}\left[\sum_{j=1}^m\bm w(k_{t,j})c_{k_{t,j}}^2\bm e_t(k_{t,j})^2\right]\\
        &\leq \frac{N_I-m}{N_I}\|\bs{e}_t\|_{\bm w}^2+\frac{mc^2}{N_I}\|\bs{e}_t\|_{\bm w}^2\\
    &\leq \left(1-\frac{m(1-c^2)}{N_I}\right)\|\bs{e}_t\|_{\bm w}^2,
        \end{split}
\end{equation}where the first inequality can be verified as \cref{eq: last term in randomized coordinate}. This concludes the proof of \cref{thm: randomized estimation 2}.

\subsection{Stability of FCFP and RCFP methods}
We denote the probability filtration $\mathcal{F}_j = \sigma(\bm n_t|t\leq j)$, which is defined due to the randomness of quantum noise up to time $j$. Define a characteristic function as
\begin{equation}\label{eq: characteristic ft}
\mathbb{I}_j=\begin{cases}
1,\quad\text{if }\{\bm n_t\}^{j-1}_{t=1}\subset B_{\gamma}(\bm n_*)\\
0,\quad \text{otherwise}.    
\end{cases}    
\end{equation}Let $X_j$ be a stochastic process defined as
\begin{equation}
X_j=\|\bs{e}_j\|_{\bm w}^2\mathbb{I}_{j}    
\end{equation}
We note that $\mathbb{I}_j$ is $\mathcal{F}_{j-1}$ measurable and $X_j$ is $\mathcal{F}_j$-measurable. In the following analysis, we assume that the initial guess $n_0$ is given as a deterministic vector in $B_{\gamma}(\bm n_*)$ in \cref{def: locally-contractive}, i.e., where the fixed-point function is contractive. 

By definition of $X_j$, we observe that
\begin{equation}
    \mathbb{P}\left\{\bm n_j\not\in B_{\gamma}(\bm n_*)\text{ for some }j\in[J]|\bm n_0\right\}\leq\mathbb{P}\left\{\sup_{1\leq j\leq J}X_j>\gamma^2|\bm n_0\right\}. 
\end{equation}
Define $E_{j-1}$ as the conditional expectation on the filtration $\mathcal{F}_{j-1}$ given $n_0$, then
\begin{equation}
    \mathbb{E}_{j-1}[X_j]=\mathbb{E}\left[\|\bs{e}_j\|_{\bm w}^2\middle|I_j=1,\bm n_0\right]\mathbb{P}\left\{\mathbb{I}_j=1|\bm n_0\right\}
\end{equation}
By definition of $\mathbb{I}_j$ and the technical result, we have
\begin{equation}
\begin{split}
\mathbb{E}\left[\|\bs{e}_j\|_{\bm w}^2\middle|\mathbb{I}_j=1,\bm n_0\right]\leq c^2\|\bs{e}_{j-1}\|_{\bm w}^2.  
\end{split}
\end{equation}
From this, we obtain that
\begin{equation}
    \mathbb{E}_{j-1}[X_j]\leq c^2\|\bs{e}_{j-1}\|_{\bm w}^2\mathbb{I}_{j-1} = c^2X_{j-1}\leq X_{j-1},
\end{equation}which yields a supermartingale,
\begin{equation}
    \mathbb{E}[X_j|\bm n_0]\leq \mathbb{E}[X_{j-1}|\bm n_0].
\end{equation}
Finally, using Markov's inequality, we arrive at
\begin{equation}
    \mathbb{P}\left\{\bm n_j\not\in B_{\gamma}(\bm n_*)\text{ for some }j\in[J]|\bm n_0\right\}\leq\mathbb{P}\left\{\sup_{1\leq j\leq J}X_j>\gamma^2|\bm n_0\right\}\leq \frac{X_0}{\gamma^2},
\end{equation}which proves the stability of FCFP.

Due to the similar property as in \cref{eq: contractive ineq3}, a similar result can be obtained for the RBCFP method as follows,
\begin{equation}
    \mathbb{P}\left\{\bm n_j\not\in B_{\gamma}(\bm n_*)\text{ for some }j\in[J]|\bm n_0\right\} \leq \frac{X_0}{\gamma^2}.
\end{equation}

\bibliographystyle{plain}

\begin{thebibliography}{10}

\bibitem{abrams1997simulation}
Daniel~S Abrams and Seth Lloyd.
\newblock Simulation of many-body fermi systems on a universal quantum
  computer.
\newblock {\em Physical Review Letters}, 79(13):2586, 1997.

\bibitem{aspuru2005simulated}
Al{\'a}n Aspuru-Guzik, Anthony~D Dutoi, Peter~J Love, and Martin Head-Gordon.
\newblock Simulated quantum computation of molecular energies.
\newblock {\em Science}, 309(5741):1704--1707, 2005.

\bibitem{avron2011randomized}
Haim Avron and Sivan Toledo.
\newblock Randomized algorithms for estimating the trace of an implicit
  symmetric positive semi-definite matrix.
\newblock {\em Journal of the ACM (JACM)}, 58(2):1--34, 2011.

\bibitem{babbush2014adiabatic}
Ryan Babbush, Peter~J Love, and Al{\'a}n Aspuru-Guzik.
\newblock Adiabatic quantum simulation of quantum chemistry.
\newblock {\em Scientific reports}, 4(1):6603, 2014.

\bibitem{babbush2018low}
Ryan Babbush, Nathan Wiebe, Jarrod McClean, James McClain, Hartmut Neven, and
  Garnet Kin-Lic Chan.
\newblock Low-depth quantum simulation of materials.
\newblock {\em Physical Review X}, 8(1):011044, 2018.

\bibitem{baer2013self}
Roi Baer, Daniel Neuhauser, and Eran Rabani.
\newblock Self-averaging stochastic kohn-sham density-functional theory.
\newblock {\em Physical review letters}, 111(10):106402, 2013.

\bibitem{baker2020density}
Thomas~E Baker and David Poulin.
\newblock Density functionals and kohn-sham potentials with minimal
  wavefunction preparations on a quantum computer.
\newblock {\em Physical Review Research}, 2(4):043238, 2020.

\bibitem{beck_real-space_2000}
Thomas~L. Beck.
\newblock Real-space mesh techniques in density-functional theory.
\newblock {\em Reviews of Modern Physics}, 72(4):1041--1080, October 2000.

\bibitem{bekas2008computation}
Constantine Bekas, Effrosini Kokiopoulou, and Yousef Saad.
\newblock Computation of large invariant subspaces using polynomial filtered
  lanczos iterations with applications in density functional theory.
\newblock {\em SIAM Journal on Matrix Analysis and Applications},
  30(1):397--418, 2008.

\bibitem{bekas2007estimator}
Costas Bekas, Effrosyni Kokiopoulou, and Yousef Saad.
\newblock An estimator for the diagonal of a matrix.
\newblock {\em Applied numerical mathematics}, 57(11-12):1214--1229, 2007.

\bibitem{bottou2018optimization}
L{\'e}on Bottou, Frank~E Curtis, and Jorge Nocedal.
\newblock Optimization methods for large-scale machine learning.
\newblock {\em SIAM review}, 60(2):223--311, 2018.

\bibitem{bowler2000efficient}
DR~Bowler and MJ~Gillan.
\newblock An efficient and robust technique for achieving self consistency in
  electronic structure calculations.
\newblock {\em Chemical Physics Letters}, 325(4):473--476, 2000.

\bibitem{brassard2000amplitude}
Gilles Brassard, Peter Hoyer, Michele Mosca, and Alain Tapp.
\newblock Amplitude amplification and quantum search algorithms.
\newblock {\em Journal of Quantum Information and Computation}, 1(4):304--320,
  2001.

\bibitem{braverman1998fast}
E~Braverman, M~Israeli, A~Averbuch, and L~Vozovoi.
\newblock A fast 3d poisson solver of arbitrary order accuracy.
\newblock {\em Journal of Computational Physics}, 144(1):109--136, 1998.

\bibitem{bressan2012lecture}
Alberto Bressan.
\newblock Lecture notes on functional analysis.
\newblock {\em Graduate studies in mathematics}, 143, 2012.

\bibitem{cances2021convergence}
Eric Canc{\`e}s, Gaspard Kemlin, and Antoine Levitt.
\newblock Convergence analysis of direct minimization and self-consistent
  iterations.
\newblock {\em SIAM Journal on Matrix Analysis and Applications},
  42(1):243--274, 2021.

\bibitem{chen2021simultaneously}
Shuai Chen, Zachary~H Aitken, Subrahmanyam Pattamatta, Zhaoxuan Wu, Zhi~Gen Yu,
  David~J Srolovitz, Peter~K Liaw, and Yong-Wei Zhang.
\newblock Simultaneously enhancing the ultimate strength and ductility of
  high-entropy alloys via short-range ordering.
\newblock {\em Nature communications}, 12(1):4953, 2021.

\bibitem{chen2023global}
Ziang Chen, Yingzhou Li, and Jianfeng Lu.
\newblock On the global convergence of randomized coordinate gradient descent
  for nonconvex optimization.
\newblock {\em SIAM Journal on Optimization}, 33(2):713--738, 2023.

\bibitem{childs2021high}
Andrew~M Childs, Jin-Peng Liu, and Aaron Ostrander.
\newblock High-precision quantum algorithms for partial differential equations.
\newblock {\em Quantum}, 5:574, 2021.

\bibitem{chow2017cyclic}
Yat~Tin Chow, Tianyu Wu, and Wotao Yin.
\newblock Cyclic coordinate-update algorithms for fixed-point problems:
  Analysis and applications.
\newblock {\em SIAM Journal on Scientific Computing}, 39(4):A1280--A1300, 2017.

\bibitem{chung1954stochastic}
K.~L. Chung.
\newblock On a stochastic approximation method.
\newblock {\em The Annals of Mathematical Statistics}, pages 463--483, 1954.

\bibitem{cleri1993tight}
Fabrizio Cleri and Vittorio Rosato.
\newblock Tight-binding potentials for transition metals and alloys.
\newblock {\em Physical Review B}, 48(1):22, 1993.

\bibitem{combettes2015stochastic}
Patrick~L Combettes and Jean-Christophe Pesquet.
\newblock Stochastic quasi-fej{\'e}r block-coordinate fixed point iterations
  with random sweeping.
\newblock {\em SIAM Journal on Optimization}, 25(2):1221--1248, 2015.

\bibitem{cui2023quantum}
Lingxia Cui, Zongmin Wu, and Hua Xiang.
\newblock Quantum radial basis function method for the poisson equation.
\newblock {\em Journal of Physics A: Mathematical and Theoretical},
  56(22):225303, 2023.

\bibitem{cytter2018stochastic}
Yael Cytter, Eran Rabani, Daniel Neuhauser, and Roi Baer.
\newblock Stochastic density functional theory at finite temperatures.
\newblock {\em Physical Review B}, 97(11):115207, 2018.

\bibitem{elstner2000self}
Marcus Elstner, Th~Frauenheim, E~Kaxiras, G~Seifert, and S~Suhai.
\newblock A self-consistent charge density-functional based tight-binding
  scheme for large biomolecules.
\newblock {\em physica status solidi (b)}, 217(1):357--376, 2000.

\bibitem{gaitan2009density}
Frank Gaitan and Franco Nori.
\newblock Density functional theory and quantum computation.
\newblock {\em Physical Review B}, 79(20):205117, 2009.

\bibitem{gale2011siesta}
Julian Gale.
\newblock Siesta: A linear-scaling method for density functional calculations.
\newblock In {\em Computational Methods for Large Systems-Electronic Structure
  Approaches for Biotechnology and Nanotechnology}, pages 45--75. Wiley \& Sons
  Inc., 2011.

\bibitem{garcia2007sub}
C.~J. Garc{\'\i}a-Cervera, J.~Lu, and W.~E.
\newblock A sub-linear scaling algorithm for computing the electronic structure
  of materials.
\newblock {\em Communications in Mathematical Sciences}, 5(4):999--1026, 2007.

\bibitem{gavini2007quasi}
Vikram Gavini, Kaushik Bhattacharya, and Michael Ortiz.
\newblock Quasi-continuum orbital-free density-functional theory: A route to
  multi-million atom non-periodic dft calculation.
\newblock {\em Journal of the Mechanics and Physics of Solids}, 55(4):697--718,
  2007.

\bibitem{ghadimi2013stochastic}
Saeed Ghadimi and Guanghui Lan.
\newblock Stochastic first-and zeroth-order methods for nonconvex stochastic
  programming.
\newblock {\em SIAM Journal on Optimization}, 23(4):2341--2368, 2013.

\bibitem{ghosh2017sparc}
Swarnava Ghosh and Phanish Suryanarayana.
\newblock Sparc: Accurate and efficient finite-difference formulation and
  parallel implementation of density functional theory: Isolated clusters.
\newblock {\em Computer Physics Communications}, 212:189--204, 2017.

\bibitem{GSLW19}
Andr{\'a}s Gily{\'e}n, Yuan Su, Guang~Hao Low, and Nathan Wiebe.
\newblock Quantum singular value transformation and beyond: exponential
  improvements for quantum matrix arithmetics.
\newblock In {\em Proceedings of the 51st Annual ACM SIGACT Symposium on Theory
  of Computing}, pages 193--204. ACM, 2019.

\bibitem{goedecker1999linear}
S.~Goedecker.
\newblock Linear scaling electronic structure methods.
\newblock {\em Reviews of Modern Physics}, 71(4):1085, 1999.

\bibitem{hafner2008ab}
J{\"u}rgen Hafner.
\newblock Ab-initio simulations of materials using vasp: Density-functional
  theory and beyond.
\newblock {\em Journal of computational chemistry}, 29(13):2044--2078, 2008.

\bibitem{hallman2022multilevel}
Eric Hallman and Devon Troester.
\newblock A multilevel approach to stochastic trace estimation.
\newblock {\em Linear Algebra and its Applications}, 638:125--149, 2022.

\bibitem{hastings2014improving}
Matthew~B Hastings, Dave Wecker, Bela Bauer, and Matthias Troyer.
\newblock Improving quantum algorithms for quantum chemistry.
\newblock {\em arXiv preprint arXiv:1403.1539}, 2014.

\bibitem{hautier2010finding}
Geoffroy Hautier, Christopher~C Fischer, Anubhav Jain, Tim Mueller, and
  Gerbrand Ceder.
\newblock Finding nature’s missing ternary oxide compounds using machine
  learning and density functional theory.
\newblock {\em Chemistry of Materials}, 22(12):3762--3767, 2010.

\bibitem{haydock1972electronic}
R~Haydock, Volker Heine, and MJ~Kelly.
\newblock Electronic structure based on the local atomic environment for
  tight-binding bands.
\newblock {\em Journal of Physics C: Solid State Physics}, 5(20):2845, 1972.

\bibitem{hohenberg1964inhomogeneous}
P.~Hohenberg and W.~Kohn.
\newblock Inhomogeneous electron gas.
\newblock {\em Physical Review}, 136(3B):B864, 1964.

\bibitem{HMW+21}
William~J Huggins, Kianna Wan, Jarrod McClean, Thomas~E O'Brien, Nathan Wiebe,
  and Ryan Babbush.
\newblock Nearly optimal quantum algorithm for estimating multiple expectation
  values.
\newblock {\em arXiv preprint arXiv:2111.09283}, 2021.

\bibitem{iiduka2019stochastic}
Hideaki Iiduka.
\newblock Stochastic fixed point optimization algorithm for classifier
  ensemble.
\newblock {\em IEEE Transactions on Cybernetics}, 50(10):4370--4380, 2019.

\bibitem{karimireddy2019efficient}
Sai~Praneeth Karimireddy, Anastasia Koloskova, Sebastian~U Stich, and Martin
  Jaggi.
\newblock Efficient greedy coordinate descent for composite problems.
\newblock In {\em The 22nd International Conference on Artificial Intelligence
  and Statistics}, pages 2887--2896. PMLR, 2019.

\bibitem{ke2007role}
San-Huang Ke, Harold~U Baranger, and Weitao Yang.
\newblock Role of the exchange-correlation potential in ab initio electron
  transport calculations.
\newblock {\em Journal of Chemical Physics}, 126(20):201102--201102, 2007.

\bibitem{ko2023stochastic}
Taehee Ko and Xiantao Li.
\newblock Stochastic algorithms for self-consistent calculations of electronic
  structures.
\newblock {\em Mathematics of Computation}, 92(342):1693--1728, 2023.

\bibitem{Kohn1965}
W.~Kohn and L.~J. Sham.
\newblock {Self-consistent equations including exchange and correlation
  effects}.
\newblock {\em Physical Review}, 140(4A):A1133--A1138, 1965.

\bibitem{li2016thick}
Ruipeng Li, Yuanzhe Xi, Eugene Vecharynski, Chao Yang, and Yousef Saad.
\newblock A thick-restart lanczos algorithm with polynomial filtering for
  hermitian eigenvalue problems.
\newblock {\em SIAM Journal on Scientific Computing}, 38(4):A2512--A2534, 2016.

\bibitem{lin2017randomized}
Lin Lin.
\newblock Randomized estimation of spectral densities of large matrices made
  accurate.
\newblock {\em Numerische Mathematik}, 136:183--213, 2017.

\bibitem{lin2019mathematical}
Lin Lin and Jianfeng Lu.
\newblock {\em A mathematical introduction to electronic structure theory}.
\newblock SIAM, 2019.

\bibitem{lin2019numerical}
Lin Lin, Jianfeng Lu, and Lexing Ying.
\newblock Numerical methods for kohn--sham density functional theory.
\newblock {\em Acta Numerica}, 28:405--539, 2019.

\bibitem{lin2016approximating}
Lin Lin, Yousef Saad, and Chao Yang.
\newblock Approximating spectral densities of large matrices.
\newblock {\em SIAM review}, 58(1):34--65, 2016.

\bibitem{lin2020optimal}
Lin Lin and Yu~Tong.
\newblock Optimal polynomial based quantum eigenstate filtering with
  application to solving quantum linear systems.
\newblock {\em Quantum}, 4:361, 2020.

\bibitem{lin2022heisenberg}
Lin Lin and Yu~Tong.
\newblock Heisenberg-limited ground-state energy estimation for early
  fault-tolerant quantum computers.
\newblock {\em PRX Quantum}, 3(1):010318, 2022.

\bibitem{lin2013elliptic}
Lin Lin and Chao Yang.
\newblock Elliptic preconditioner for accelerating the self-consistent field
  iteration in kohn--sham density functional theory.
\newblock {\em SIAM Journal on Scientific Computing}, 35(5):S277--S298, 2013.

\bibitem{lin2014accelerated}
Qihang Lin, Zhaosong Lu, and Lin Xiao.
\newblock An accelerated proximal coordinate gradient method.
\newblock {\em Advances in Neural Information Processing Systems}, 27, 2014.

\bibitem{liou2020scalable}
Kai-Hsin Liou, Chao Yang, and James~R Chelikowsky.
\newblock Scalable implementation of polynomial filtering for density
  functional theory calculation in parsec.
\newblock {\em Computer Physics Communications}, 254:107330, 2020.

\bibitem{marques2012libxc}
Miguel~AL Marques, Micael~JT Oliveira, and Tobias Burnus.
\newblock Libxc: A library of exchange and correlation functionals for density
  functional theory.
\newblock {\em Computer physics communications}, 183(10):2272--2281, 2012.

\bibitem{martin2004@book}
R.~M. Martin.
\newblock {\em {Electronic Structure: Basic Theory and Practical Methods}}.
\newblock {Cambridge University Press}, 2011.

\bibitem{merrick1995multigrid}
Michael~P Merrick, Karthik~A Iyer, and Thomas~L Beck.
\newblock Multigrid method for electrostatic computations in numerical density
  functional theory.
\newblock {\em The Journal of Physical Chemistry}, 99(33):12478--12482, 1995.

\bibitem{meyer2021hutch++}
Raphael~A Meyer, Cameron Musco, Christopher Musco, and David~P Woodruff.
\newblock Hutch++: Optimal stochastic trace estimation.
\newblock In {\em Symposium on Simplicity in Algorithms (SOSA)}, pages
  142--155. SIAM, 2021.

\bibitem{motamarri2020dft}
Phani Motamarri, Sambit Das, Shiva Rudraraju, Krishnendu Ghosh, Denis Davydov,
  and Vikram Gavini.
\newblock Dft-fe--a massively parallel adaptive finite-element code for
  large-scale density functional theory calculations.
\newblock {\em Computer Physics Communications}, 246:106853, 2020.

\bibitem{nemirovski2009robust}
Arkadi Nemirovski, Anatoli Juditsky, Guanghui Lan, and Alexander Shapiro.
\newblock Robust stochastic approximation approach to stochastic programming.
\newblock {\em SIAM Journal on optimization}, 19(4):1574--1609, 2009.

\bibitem{nesterov2012efficiency}
Yu~Nesterov.
\newblock Efficiency of coordinate descent methods on huge-scale optimization
  problems.
\newblock {\em SIAM Journal on Optimization}, 22(2):341--362, 2012.

\bibitem{nesterov2017efficiency}
Yurii Nesterov and Sebastian~U Stich.
\newblock Efficiency of the accelerated coordinate descent method on structured
  optimization problems.
\newblock {\em SIAM Journal on Optimization}, 27(1):110--123, 2017.

\bibitem{nielsen2011quantum}
Michael~A Nielsen and Isaac~L Chuang.
\newblock {\em Quantum Computation and Quantum Information}.
\newblock Cambridge University Press, 2011.

\bibitem{nutini2015coordinate}
Julie Nutini, Mark Schmidt, Issam Laradji, Michael Friedlander, and Hoyt
  Koepke.
\newblock Coordinate descent converges faster with the gauss-southwell rule
  than random selection.
\newblock In {\em International Conference on Machine Learning}, pages
  1632--1641. PMLR, 2015.

\bibitem{o2016scalable}
Peter~JJ O’Malley, Ryan Babbush, Ian~D Kivlichan, Jonathan Romero, Jarrod~R
  McClean, Rami Barends, Julian Kelly, Pedram Roushan, Andrew Tranter, Nan
  Ding, et~al.
\newblock Scalable quantum simulation of molecular energies.
\newblock {\em Physical Review X}, 6(3):031007, 2016.

\bibitem{parr1995density}
R.~G. Parr and W.~Yang.
\newblock {\em Density-functional theory of atoms and molecules}.
\newblock Oxford University Press, 1995.

\bibitem{peng2016arock}
Zhimin Peng, Yangyang Xu, Ming Yan, and Wotao Yin.
\newblock Arock: an algorithmic framework for asynchronous parallel coordinate
  updates.
\newblock {\em SIAM Journal on Scientific Computing}, 38(5):A2851--A2879, 2016.

\bibitem{perdew1996generalized}
John~P Perdew, Kieron Burke, and Yue Wang.
\newblock Generalized gradient approximation for the exchange-correlation hole
  of a many-electron system.
\newblock {\em Physical review B}, 54(23):16533, 1996.

\bibitem{perdew1992accurate}
John~P Perdew and Yue Wang.
\newblock Accurate and simple analytic representation of the electron-gas
  correlation energy.
\newblock {\em Physical review B}, 45(23):13244, 1992.

\bibitem{perdew1981local}
John~P. Perdew and Alex Zunger.
\newblock Local density-functional theory and its application to atoms and
  molecules.
\newblock {\em Physical Review B}, 23(10):5048--5079, 1981.

\bibitem{persson2022improved}
David Persson, Alice Cortinovis, and Daniel Kressner.
\newblock Improved variants of the hutch++ algorithm for trace estimation.
\newblock {\em SIAM Journal on Matrix Analysis and Applications},
  43(3):1162--1185, 2022.

\bibitem{rall2020quantum}
Patrick Rall.
\newblock Quantum algorithms for estimating physical quantities using block
  encodings.
\newblock {\em Physical Review A}, 102(2):022408, 2020.

\bibitem{robbins1951stochastic}
H.~Robbins and S.~Monro.
\newblock A stochastic approximation method.
\newblock {\em The Annals of Mathematical Statistics}, pages 400--407, 1951.

\bibitem{saha2013nonasymptotic}
Ankan Saha and Ambuj Tewari.
\newblock On the nonasymptotic convergence of cyclic coordinate descent
  methods.
\newblock {\em SIAM Journal on Optimization}, 23(1):576--601, 2013.

\bibitem{schofield2012spectrum}
Grady Schofield, James~R Chelikowsky, and Yousef Saad.
\newblock A spectrum slicing method for the kohn--sham problem.
\newblock {\em Computer Physics Communications}, 183(3):497--505, 2012.

\bibitem{seifert2012density}
Gotthard Seifert and Jan-Ole Joswig.
\newblock Density-functional tight binding—an approximate density-functional
  theory method.
\newblock {\em Wiley Interdisciplinary Reviews: Computational Molecular
  Science}, 2(3):456--465, 2012.

\bibitem{senjean2023toward}
Bruno Senjean, Saad Yalouz, and Matthieu Sauban{\`e}re.
\newblock Toward density functional theory on quantum computers?
\newblock {\em SciPost Physics}, 14(3):055, 2023.

\bibitem{sharma2018calculation}
Abhiraj Sharma and Phanish Suryanarayana.
\newblock On the calculation of the stress tensor in real-space kohn-sham
  density functional theory.
\newblock {\em The Journal of chemical physics}, 149(19):194104, 2018.

\bibitem{soler2002siesta}
J.~M. Soler, E.~Artacho, J.~D. Gale, A.~Garc{\'\i}a, J.~Junquera,
  P.~Ordej{\'o}n, and D.~S{\'a}nchez-Portal.
\newblock The {SIESTA} method for ab initio order-{N} materials simulation.
\newblock {\em Journal of Physics: Condensed Matter}, 14(11):2745, 2002.

\bibitem{tesch2001applying}
Carmen~M Tesch, Lukas Kurtz, and Regina de~Vivie-Riedle.
\newblock Applying optimal control theory for elements of quantum computation
  in molecular systems.
\newblock {\em Chemical Physics Letters}, 343(5-6):633--641, 2001.

\bibitem{toth_local_2017}
A.~Toth, J.~A. Ellis, T.~Evans, S.~Hamilton, C.~T. Kelley, R.~Pawlowski, and
  S.~Slattery.
\newblock Local {Improvement} {Results} for {Anderson} {Acceleration} with
  {Inaccurate} {Function} {Evaluations}.
\newblock {\em SIAM Journal on Scientific Computing}, 39(5):S47--S65, January
  2017.

\bibitem{toth_convergence_2015}
A.~Toth and C.~T. Kelley.
\newblock Convergence {Analysis} for {Anderson} {Acceleration}.
\newblock {\em SIAM Journal on Numerical Analysis}, 53(2):805--819, January
  2015.

\bibitem{trefethen2019approximation}
Lloyd~N Trefethen.
\newblock {\em Approximation Theory and Approximation Practice, Extended
  Edition}.
\newblock SIAM, 2019.

\bibitem{tsitsiklis1994asynchronous}
John~N Tsitsiklis.
\newblock Asynchronous stochastic approximation and q-learning.
\newblock {\em Machine learning}, 16:185--202, 1994.

\bibitem{vecharynski2015projected}
Eugene Vecharynski, Chao Yang, and John~E Pask.
\newblock A projected preconditioned conjugate gradient algorithm for computing
  many extreme eigenpairs of a hermitian matrix.
\newblock {\em Journal of Computational Physics}, 290:73--89, 2015.

\bibitem{wen2016trace}
Zaiwen Wen, Chao Yang, Xin Liu, and Yin Zhang.
\newblock Trace-penalty minimization for large-scale eigenspace computation.
\newblock {\em Journal of Scientific Computing}, 66:1175--1203, 2016.

\bibitem{wolfowitz1952stochastic}
J.~Wolfowitz.
\newblock On the stochastic approximation method of {Robbins and Monro}.
\newblock {\em The Annals of Mathematical Statistics}, 23(3):457--461, 1952.

\bibitem{wright2015coordinate}
Stephen~J Wright.
\newblock Coordinate descent algorithms.
\newblock {\em Mathematical programming}, 151(1):3--34, 2015.

\bibitem{yang2009kssolv}
Chao Yang, Juan~C Meza, Byounghak Lee, and Lin-Wang Wang.
\newblock Kssolv—a matlab toolbox for solving the kohn-sham equations.
\newblock {\em ACM Transactions on Mathematical Software (TOMS)}, 36(2):1--35,
  2009.

\bibitem{yoo2019atomic}
Hyobin Yoo, Rebecca Engelke, Stephen Carr, Shiang Fang, Kuan Zhang, Paul
  Cazeaux, Suk~Hyun Sung, Robert Hovden, Adam~W Tsen, Takashi Taniguchi, et~al.
\newblock Atomic and electronic reconstruction at the van der waals interface
  in twisted bilayer graphene.
\newblock {\em Nature materials}, 18(5):448--453, 2019.

\bibitem{zhou2006self}
Yunkai Zhou, Yousef Saad, Murilo~L Tiago, and James~R Chelikowsky.
\newblock Self-consistent-field calculations using chebyshev-filtered subspace
  iteration.
\newblock {\em Journal of Computational Physics}, 219(1):172--184, 2006.

\end{thebibliography}

\end{document}